\documentstyle[amssymb,preprint,aps]{revtex}
\tightenlines
\begin{document}
\draft
\title{
The metal-insulator transition in amorphous Si$_{1-x}$Ni$_x$:
\linebreak Evidence for Mott's minimum metallic conductivity
}
\author{A.~M\"obius,$^1$\footnote{e-mail: a.moebius@ifw-dresden.de}
C.~Frenzel,$^1$ R.~Thielsch,$^{1,2}$ R.~Rosenbaum,$^3$ 
C.J.~Adkins,$^4$ M.~Schreiber,$^5$ H.-D.~Bauer,$^1$ 
R.~Gr\"otzschel,$^6$ V.~Hoffmann,$^1$ 
T.~Krieg,$^1$ N.~Matz,$^1$ H.~Vinzelberg,$^1$ and M.~Witcomb$^7$
}
\address{
$^1$Institute for Solid State and Materials Research Dresden,
D-01171 Dresden, Germany,\\
$^2$Fraunhofer Institute for Applied Optics and Precision Mechanics, 
Schillerstr.\ 1, D-07745 Jena, Germany,\\
$^3$Tel Aviv University, School of Physics and Astronomy,
Raymond and Beverly Sackler Faculty of Exact Sciences, Ramat Aviv, 
69978, Israel,\\
$^4$Cavendish Laboratory, Madingley Road, Cambridge CB3 0HE, UK,\\
$^5$Technical University Chemnitz, Institute of Physics, 
D-09107 Chemnitz, Germany,\\
$^6$Research Center Rossendorf, Institute for Ion Beam Physics and 
Materials Research, D-01314 Dresden, Germany,\\
$^7$University of the Witwatersrand, Electron Microscope Unit, Private
Bag 3, Wits, 2050, South Africa
}
\date{\today}
\maketitle

\begin{abstract}
We study the metal-insulator transition in two sets of amorphous
Si$_{1-x}$Ni$_x$ films. The sets were prepared by different, 
electron-beam-evaporation-based technologies: evaporation of the alloy, 
and gradient deposition from separate Ni and Si crucibles. The 
characterization included electron and scanning tunneling microscopy, 
glow discharge optical emission spectroscopy, energy dispersive X-ray 
analysis, and Rutherford back scattering. Investigating the logarithmic 
temperature derivative of the conductivity, 
$w = \mbox{d} \ln \sigma / \mbox{d} \ln T$, we observe that, for 
insulating samples, $w(T)$ shows a minimum, increasing at both low and 
high $T$. Both the minimum value of $w$ and the corresponding 
temperature seem to tend to zero as the transition is approached. The 
analysis of this feature of $w(T,x)$ leads to the conclusion that the 
transition in Si$_{1-x}$Ni$_x$ is very likely discontinuous at zero 
temperature in agreement with Mott's original views.
\end{abstract}


\narrowtext

\section{Introduction}

Metal-insulator transitions (MIT) in disordered systems have attracted
much interest, theoretical as well as experimental, for forty years. 
Milestones on this way were the concepts of Anderson localization
\cite{Ande}, of minimum metallic conductivity \cite{Mott.72}, the 
scaling theory of localization \cite{Abra.etal}, and the renormalization 
group approach incorporating the electron-electron interaction into 
localization theory \cite{Fink.83a,Fink.83b,Fink.84}; for surveys 
see Refs.\ 
\onlinecite{Lee.Rama,Moe.85,Mott.90,Kram.McKi,Beli.Kirk,Edwa.95}. 

Nevertheless, important questions have not been finally settled yet: 
Contrary to Mott's prediction of a finite minimum metallic conductivity, 
the scaling theory of localization predicts continuity of the 
conductivity at the MIT for three-dimensional systems. That means, for 
the transition being approached from the metallic side at zero 
temperature, $T = 0$, the conductivity, $\sigma(T,x)$, is expected to 
vanish continuously according to 
\begin{equation}
\sigma(0,x) \propto |x-x_{\rm c}|^\nu \,, 
\end{equation}
where $x$ stands for the control parameter (concentration / magnetic 
field / stress ...), and $x_{\rm c}$ for its value at the MIT. The value 
of the critical exponent $\nu$ has been a subject of controversial 
debate since its first measurements, see e.g.\ Refs.\
\onlinecite{Paal.etal,Thom.etal.83,Hert.etal,Peih.etal,Stup.etal,Cast}. 
However, several studies have pointed to inconsistencies of the 
zero-temperature extrapolations inherent in the critical-exponent 
determination 
\cite{Moe.85,Edwa.95,Cast,Moe.89,Hirs.etal,Thom.Holc,Rose.etal.94}.
These problems could, of course, arise from quantitative failure of the
extrapolation formulae used or from imperfections of samples or $\sigma$ 
and $T$ measurements, but they could also be caused by discontinuity of 
$\sigma(0,x)$ at $x_{\rm c}$ \cite{Moe.85,Edwa.95}, in which case the 
critical-exponent puzzle would simply be an artifact arising from 
unphysical fitting. The unclear situation described is the main 
motivation of the present study.

The recent discovery of a sharp MIT in two-dimensional disordered 
systems \cite{Krav.etal.94,Krav.etal.95,Krav.etal.96,Lubk} has drawn 
interest to this subject, too. The scaling theory of localization of 
non-interacting electrons \cite{Abra.etal} states that two-dimensional 
systems exhibit principally only activated conduction, and denies the 
existence of an MIT in this situation. Hence it does not describe nature 
in this case. This failure is a strong additional motivation to 
re-consider the applicability of that theory to the MIT of 
three-dimensional systems.

In greater detail, the critical step of most of the recent studies of 
the MIT in three-dimensional disordered systems is the zero-temperature 
extrapolation, that is the determination of the limit 
$\sigma(T \rightarrow 0,x)$, for samples close to the MIT. This 
extrapolation can be based either on a microscopic theory, or on an 
empirically found relation supposed to be valid down to $T = 0$. 

The extrapolation is particularly difficult if a strong, but 
non-exponential $T$ dependence of $\sigma$ is observed. In such a case, 
two interpretations are possible:\\
(i) The sample could be metallic. This situation is mostly analyzed 
in terms of theories by Altshuler and Aronov \cite{Alts.Aron}, or 
Newson and Pepper \cite{News.Pepp}, which yield 
\begin{equation}
\sigma(T,x) = a(x) + b(x) \cdot T^p
\end{equation}
with $p = 1/2$, and $1/3$, respectively. The former theory models the 
superposition of electron-electron interaction and disorder, but it is 
a perturbation theory so that its applicability very close to the 
transition is at least questionable. The latter theory considers the
$T$-dependent drop of the diffusion constant as the decisive variation, 
and yields a power law with exponent 1/3 for $x = x_{\rm c}$.\\
(ii) The sample could exhibit activated conduction with the 
characteristic energy being smaller than the lowest experimentally 
accessible $T$. Thus it would have to be classified as insulating (at 
$T = 0$). If, as generally taken for granted, the characteristic energy 
vanishes continuously as $x \rightarrow x_{\rm c}$, such an $x$ region 
exists always, see Fig.\ 1. But, to the best of our knowledge, there is 
no appropriate microscopic theory for a related quantitative analysis 
available.

Unfortunately, when considering non-exponential 
$\sigma(T,x = {\rm const.})$ dependences, many of the recent 
experimental studies analyze the data only in terms of (i). Moreover, to 
agree with Eq.\ (1), they presume that, as $x \rightarrow x_{\rm c}$, 
the parameter $a$ tends to 0. So it is not surprising that, when 
checking the applicability of Eq.\ (2) by considering the logarithmic 
derivative $\mbox{d} \ln \sigma / \mbox{d} \ln T$, doubts arose in 
several cases \cite{Moe.89,Hirs.etal,Thom.Holc,Rose.etal.94}: this 
quantity must vanish as $T \rightarrow 0$ for metallic samples, but, 
instead, often increases or stays approximately constant for the 
sample(s) assumed to be particularly close to the MIT. Therefore one 
should determine the MIT point by simultaneously considering both sides 
of the transition. Moreover, adjusting free parameters should be avoided 
to the largest possible extent. Such an approach was successful for 
amorphous Si$_{1-x}$Cr$_x$, see Refs.\
\onlinecite{Moe.etal.83,Moe.etal.85,Moe.87,Moe.90a,Moe.90b}. 
A simple phenomenological model of $\sigma(T,x)$ describing both sides 
of the MIT was constructed for that system starting from the detection
of universal features in $\sigma(T,x)$, in particular from a scaling law
for the $T$ dependence in the activated region, 
$\sigma(T,x) = \sigma(T/T_0(x))$. However, annealing breaks this 
universality.

Due to the controversial situation described above, the aim of the 
present work is to analyze the $\sigma(T,x = {\rm const.})$ relations 
close to the transition in as unbiased a way as possible, without 
fitting. For that we study amorphous silicon nickel films, 
a-Si$_{1-x}$Ni$_x$, prepared by electron beam evaporation. Thus, though
the above discussion concerns a general problem occuring for an 
arbitrary type of the control parameter $x$, we focus now on a special 
case, and, in the following, $x$ stands only for metal concentration in 
an amorphous alloy. In our study, we put emphasis on two methodical 
points: Firstly, we attach particular importance to the accuracy of the 
$\mbox{d} \ln \sigma / \mbox{d} \ln T$ values. Secondly, in order to 
avoid being mislead by having used special preparation conditions, we 
investigate two series of samples, prepared by groups in Dresden and Tel 
Aviv.

There were several reasons for us to select a-Si$_{1-x}$Ni$_x$ for a
detailed investigation:\\ 
-- We were looking for an alloy which is sufficiently stable but 
differs substantially from a-Si$_{1-x}$Cr$_x$ which we had studied in 
detail previously
\cite{Moe.etal.83,Moe.etal.85,Moe.87,Moe.90a,Moe.90b}.
According to the melting temperatures \cite{Phasediag}, 
a-Si$_{1-x}$Ni$_x$ can be expected to be rather stable, at least far 
more than a-Si$_{1-x}$Au$_x$ \cite{Nish.etal}, or a-Ge$_{1-x}$Au$_x$ 
\cite{Dods.etal}. On the other hand, there is a deep eutectic point 
close to the MIT for the system Si-Cr, but not for Si-Ni 
\cite{Phasediag}. Therefore, in a quenching experiment, it should be far 
more difficult to realize an amorphous structure close to the MIT in 
Si$_{1-x}$Ni$_x$ than in Si$_{1-x}$Cr$_x$ \cite{Ohri.glass}. However, it 
is an open question, to what extent such considerations can be applied 
also to films produced by evaporation, or by sputtering.\\ 
-- A second important difference between a-Si$_{1-x}$Ni$_x$ and
a-Si$_{1-x}$Cr$_x$ concerns the character of the silicide which should 
be formed first in annealing the samples, cf.\ Refs.\ 
\onlinecite{Olow.etal,Colg.etal,Sobe.Zies}: Si$_2$Ni is a metal, whereas 
Si$_2$Cr is a semiconductor with an indirect gap of roughly 0.3~eV 
\cite{metal_silic}.\\
-- Conductivity studies (partly taking magnetoresistance into account) on 
hydrogenated and unhydrogenated sputtered a-Si$_{1-x}$Ni$_x$ films 
\cite{Abke.etal.92a,Abke.etal.92b,Damm.etal,Damm}, as well as on
polycrystalline films \cite{Coll1,Coll2,Sega.etal}, and on the influence 
of annealing \cite{Sega.etal,Belu.etal} were available for comparison in 
the literature. Moreover, in the vicinity of the MIT, the electronic 
structure of a-Si$_{1-x}$Ni$_x$ was recently studied by XPS and UPS
\cite{Isob.etal}. For a-Si$_{1-x}$Ni$_x$:H, a structure study
\cite{Edwa.etal}, optical conductivity measurements \cite{Davi.etal}, 
also under pressure \cite{Asal.etal}, and electronic structure 
investigations by XPS and XES \cite{Gheo.etal} were reported in the 
literature. Last but not least, some of the authors have gained detailed 
experience in trying fits of $\sigma(T)$ by conventional theories for 
a-Si$_{1-x}$Ni$_x$ samples prepared by electron-beam evaporation in Tel 
Aviv \cite{Rose.etal.97a,Rose.etal.97b}.

The present paper is organized as follows: Section II describes the
sample preparation. Section III is devoted to the structural and 
chemical characterization, including an STM investigation. Section IV 
gives some experimental details of the conductivity measurements, 
whereas Section V analyses them in detail, and lists common and 
differing features of the two conductivity data sets. In Section VI, we 
discuss the qualitative character of the MIT. Finally, in Section VII, 
the conclusions from our findings are summarized.

\section{Film preparation} 

We compare a-Si$_{1-x}$Ni$_x$ films, prepared by two different 
technologies. Both technologies are based on electron-beam evaporation, 
but they have specific advantages and disadvantages:\\
(A) Direct evaporation of SiNi alloys of various compositions from one
crucible was used in Dresden, cf.\ Ref.\ \onlinecite{Moe.etal.83}. In 
this case, changes of power cause changes of the composition of the
films by modifying the vapor pressure ratio of the alloy components via
the changed temperature of the ingot. Hence, the deposition rate has 
carefully to be held constant. Moreover, it is essential that the whole 
ingot is liquid; a check for contamination from the copper crucible 
showed it to be negligible. The $x$ drift over the film thickness 
arising from Ni enrichment in the ingot material during evaporation is 
very small, see section III, since only a small portion of the ingot was 
evaporated in each cycle. Moreover, this drift is far smaller than the 
$x$ difference between samples prepared in successive cycles using the 
same ingot, because the time which we waited until rate and pressure 
were stable is considerably larger than the deposition duration. Typical 
conditions were as follows: residual gas pressure during deposition 
$\approx 4 \cdot 10^{-6}\ {\rm mbar}$, and rate 
$\approx 2\ {\rm nm/s}$.\\
(B) Co-evaporation of Ni and Si from separate crucibles was used in 
Tel Aviv to prepare in one deposition a whole set of samples. This 
method uses the varying incidence angles and distances to the sources to 
produce a composition gradient along a substrate. The film is then cut 
into narrow samples with their axes (the direction of current flow) 
perpendicular to the composition gradient \cite{Rose.etal.97a}. This 
technique has two advantages: Composition determination is related to a 
geometrical measurement, and neighboring samples are prepared under 
almost identical conditions. Thus the random errors in studying $x$ 
dependences can considerably be reduced. However, the crucial point is 
that one has to take great care to hold the ratio of the evaporation 
rates of both crucibles constant. The conditions were as follows: 
residual gas pressure during deposition 
$\approx 1 \cdot 10^{-5}\ {\rm mbar}$, and, for the samples close to the 
MIT, rate $\approx 0.8\ {\rm nm/s}$; for more details see Ref.\ 
\onlinecite{Rose.etal.97a}.

In the following, we refer to the samples prepared by technology A
by the numbers 1 -- 6, and to the samples produced by method B by the 
letters a -- j. These two sample sets are labeled as A and B according 
to the preparation. Tables I and II give $x$, film thickness, and 
conductivity at 300 and 4.2~K for the sets A and B, respectively. 
Samples 6 and i, both deep in the metallic region, and sample j, close 
to the MIT, are only included for comparison of $x$ scales.

\section{Chemical and structural characterization}

\subsection{Composition}

The chemical composition of the samples of set A was determined by 
Rutherford backscattering (RBS) at the Research Center Rossendorf within 
this work. For that we used Sigradur substrates (glassy carbon). We 
confirmed the consistency of these data by glow discharge optical 
emission spectroscopy (GDOES). Energy dispersive X-ray analysis (EDX) 
by means of a LINK AN10000 EDS system attached to a JSM-840 SEM
was used in determining the composition for sample set B in 
Witwatersrand, South Africa, within Refs.\ 
\onlinecite{Rose.etal.97a,Rose.etal.97b}. The results of these analyses 
are given in Tables I and II.

In order to compare both $x$ scales, we performed additional
EDX analyses for samples 3, 6, f, h, and i by means of a Tracor/Noran
spectrometer Voyager IIa, attached to a transmission electron microscope 
(TEM) Philips CM20FEG, in Dresden. For that, pieces of the film were 
scraped from the substrate with a steel knife, and placed on a 
carbon-film-coated copper microscope grid. Moreover, for sample j, 
the Witwatersrand EDX result was checked by an RBS analysis performed
in Faure, South Africa, within Ref.\ \onlinecite{Rose.etal.97a}.

The results of these analyses are presented in Table III. It illustrates 
the inconsistencies in the $x$ scales which were already pointed out in 
Ref.\ \onlinecite{Rose.etal.97a}. Note that the Dresden EDX data 
underestimate our Rossendorf RBS values by roughly 5~at$\%$ Ni for 
sample 3 as well as for sample 6. Moreover, comparing the Dresden EDX 
data with the Witwatersrand EDX values \cite{Rose.etal.97a}, we conlcude 
that the latter overestimate the Ni content, according to our RBS, by 
roughly 6~at$\%$ close to the MIT. This is confirmed by the comparison 
of South African EDX and RBS data for sample j, yielding a difference of 
5.3~at$\%$. For sample i, the EDX value of Ref.\ 
\onlinecite{Rose.etal.97a} might be very close to the Ni content 
according to our RBS scale -- Table III suggests that the systematic 
deviation of the Dresden EDX values from our RBS data could be roughly 
constant. Finally, an additional check with an SiNi$_2$ bulk standard 
showed the Witwatersrand EDX to overestimate the true value by roughly 
3~at$\%$ there. 

We consider the RBS as likely to give the most accurate $x$ values 
since this method is quantitative from first principles and does not 
require elemental standards \cite{Ohri}. This is confirmed by the
good agreement of both RBS scales in Table III. In our case, the 
relative uncertainty of the corresponding Ni:Si ratios was estimated to 
be better than 2~$\%$, so that the related absolute error of the Ni 
content should not exceed 0.3~at$\%$ in the MIT region. The quantitative 
result of the EDX analysis however is influenced by a series of effects, 
in particular this method needs standards and various corrections 
\cite{EDX}. Therefore, the systematic errors can amount to several 
at$\%$ so that the differences between the columns of Table III are not 
surprising.

The $x$ data, given in Tables I to III, and discussed in the 
previous paragraphs, are obtained from the Ni:Si ratio ignoring all 
kinds of contamination. Different kinds of analysis showed oxygen 
contamination to be the most relevant one. By means of EDX, we obtained 
4.4 and 11~at$\%$ as oxygen concentration in the samples 3 and h, 
respectively. If the oxygen were distributed homogeneously in the films, 
a Ni concentration of 20~at$\%$ determined from the Ni:Si ratio would 
correspond to true 19.2 and 18~at$\%$ Ni for series A and B, 
respectively. However, some of the oxygen is expected to be concentrated 
close to the film surface and the film-substrate interface, so that this 
quantitative information should not be overvalued.

Both sets of films were found to be well homogeneous. Laterally we 
checked this by EDX and electrical measurements for both sets. Changes 
in vertical direction might be critical in both technologies. We used 
GDOES to clarify this point. Because of the non-conducting substrate, we 
applied our self-developed radiofrequency equipment, a new and very 
suitable device for studying vertical concentration variations 
\cite{Hoff}. For series B, the vertical homogeneity was also checked by 
monitoring the quotient of the Ni and Si evaporation rates during film
deposition \cite{Rose.etal.97a}. 

By means of GDOES, the Ni:Si ratio was found to deviate from the 
respective mean values through the thickness of the samples by not more 
than 5~$\%$ for both sample sets, corresponding to $\pm 0.7\ {\rm at}\%$ 
Ni close to the MIT. Fig.\ 2 shows depth profiles for series A and B. 
Note the constancy of the Ni signal. The variation of the Si signal at 
the beginning is an artifact. It arises from the strong dependence of 
the intensity of the Si line used on the changing plasma parameters at 
ignition. This variation is caused by surface contamination indicated
by the C (carbon) signal; this was checked by additional experiments. 
Therefore, we estimated the relative variation of the Ni:Si ratio from 
the Ni signal.

\subsection{Structure}

The microstructure of the samples of both sets was investigated by TEM. 
In all cases, the samples were found to be well amorphous.  However, the 
existence of crystalline regions with diameters below 2~nm cannot be 
excluded. Such a clustering is suggested by \onlinecite{Edwa.etal}, cf.\
\onlinecite{Raap.etal}.

The topography of the a-Si$_{1-x}$Ni$_x$ films was studied ex situ by 
means of a scanning tunneling microscope (STM) under UHV conditions 
($p < 1 \cdot 10^{-9} {\rm mbar}$). For that purpose, we used a modified 
Omicron-UHV-STM with etched tungsten tip and partly self-made 
electronics. A typical topography of sample 3 is given in Fig.\ 3. Three 
corresponding cross-sections are presented in Fig.\ 4. For sample f, we 
obtained very similar results.

Figs.\ 3 and 4 show that there are fluctuations in the film thickness 
of typically $\pm 2$~nm. In rare cases, the fluctuations reach an 
altitude of $\pm 7$~nm. The horizontal diameter of these hills amounts 
to the order of 30~nm. However, there is also a fine structure on a 
smaller scale, which fact might be related to some fractal structuring
\cite{Bara.Stan}. Unfortunately, we have no information to what extent 
these structures can be considered as representing a surface roughness 
only, or whether they originate from a columnar film growth, see e.g.\
Ref.\ \onlinecite{Bala.Zang}.

The surface roughness could cause differences between technologies A and 
B concerning the chemical homogeneity of the films when considering a 
mesoscopic length scale: The local chemical composition is independent 
of the surface slope in technology A. However, during gradient film 
deposition by means of technology B, the variation of the surface slope 
causes a fluctuation of the Si and Ni incidence angles, where an 
increase of the former is correlated with a decrease of the latter. Thus 
the chemical composition of these films ought to fluctuate on a 
mesoscopic length scale \cite{Kran.Lodd}. 

To estimate quantatively this effect is difficult for two reasons: (i)
The fluctuation strength depends on the size of the surface area 
taken into account in averaging. (ii) We have no information on surface 
and volume diffusion lengths limiting composition fluctuations. As an 
example, we consider a cube with edge length 2.4~nm, and assume the
diffusion lengths not to exceed this value. The typical surface slope,
estimated as root mean square of the difference quotient 
$\Delta z/\Delta x$ (with $\Delta x = 2.4\ {\rm nm}$) for several cross 
sections, corresponds to an angle of 14 degrees. For sample f, the 
incidence angles of the Si and Ni atoms, arriving from opposite sides, 
are roughly 10 and 45 degrees. This sample has a mean Ni content of 
24.8~at$\%$ according to the EDX scale of Ref.\ 
\onlinecite{Rose.etal.97a}. Thus the fluctuation of the incidence angle 
due to the surface roughness causes composition fluctuations with a mean 
deviation of $\pm 6$~at$\%$ Ni. This value is by a factor of 3 larger 
than the finite size induced random dispersion of the Ni content. 
However, since the regions of maximum and minimum Ni content do not form 
a percolative network, they have to be considered as local inclusions.

\section{Conductivity measurements}

Three types of measurements have been performed to study the $T$
dependence of $\sigma$:\\
(a) 2 -- 300 K using a self-built cryostat insert based on a bell 
  technique, which permits very accurate stabilization of $T$ over the 
  whole interval, so that a high accuracy of the values of the 
  logarithmic derivative is guaranteed,\\ 
(b) 0.45 -- 4.2 K using an RMC $^3$He cryostat, and\\
(c) 35 -- 900 mK using a dilution refrigerator Kelvinox 300 from 
Oxford Instruments.\\
Measurements (a) of the B samples are repetitions with increased
accuracy of the investigations in Refs.\
\onlinecite{Rose.etal.97a,Rose.etal.97b}.
Some of the measurements (b) were already published in Refs.\
\onlinecite{Rose.etal.97a,Rose.etal.97b}. However, within this study
they were re-analyzed, and the values of the logarithmic derivative were 
re-calculated by means of Eqs.\ (8,9) below.

Two methods of making contacts to the samples were used: The leads were 
attached either by pressing indium tabs onto the sample, or by silver 
paste. Both these technologies avoid post-preparation heating of the 
films, but they need care to avoid contact problems in thermal cycling.

A specific problem consists in the aging of the samples which can cause 
a resistivity change up to the order of 10~$\%$. Thus we performed 
annealing experiments: Metallic films are influenced only slightly by 
annealing below roughly 150$^\circ$C, but annealing up to about 
300$^\circ$C leads to great changes, cf.\ Refs.\ 
\onlinecite{Sega.etal,Belu.etal}. Repeated thermal cycling between 
cryogenic and room temperatures can cause resistivity changes by a 
few $\%$ as well, but no drift is detectable in thermal cycling between 
2 and 4.2~K. The usual aging concerns only the absolute value of the 
resistivity; the resistivity ratio $R(T)/R(300\ {\rm K})$ was observed 
to remain almost unchanged. Therefore we do not consider the sample 
after aging / thermal cycling to be a different sample, as if prepared 
under different conditions, and we simply scale the resistivity when 
necessary to retain the original value at an appropriate $T$. However, 
this scaling is irrelevant for the consideration of the logarithmic 
derivative, on which the main conclusion of this paper is based.

Overviews for both the data sets A and B are shown in Fig.\ 5. Both data 
sets exhibit consistent behavior, with some systematic differences 
between them. For comparison, Fig.\ 5a includes data of two sputtered 
a-Si$_{1-x}$Ni$_x$ films from Ref.\ \onlinecite{Isob.etal}, resembling
the curves of our evaporated samples very much. However, it is almost 
impossible to obtain reliable information on physical mechanisms 
directly from these only weakly structured curves. 

\section{Conductivity data analysis}

\subsection{Standard extrapolation}

The standard analysis for determining the critical concentration of the 
MIT, based on extrapolation according to Eq.\ (2) with $p = 1/2$, is 
presented in Fig.\ 6. Indeed, if only data points between roughly 2 
and 30~K are taken into consideration, linear relations corresponding to 
Eq.\ (2) are found in the $\sigma$ versus $T^{1/2}$ representations. 
According to this analysis the MIT should be located just below the $x$ 
of sample 1 in set A, and between samples c and d in set B. However, 
inclusion of the data around and below about 1~K shows that this 
classification is questionable.

\subsection{The logarithmic derivative}

More insight can be gained by considering the logarithmic derivative,
\begin{equation}
w(T) = \frac{\mbox{d} \ln \sigma(T)}{\mbox{d} \ln T}\,.
\end{equation}
Its limiting behavior for vanishing $T$ allows one to detect the 
transition point \cite{Moe.89}.

As a first approach to an understanding of how $w(T,x)$ reflects the 
character of the MIT, let us assume that, for a given sample, the 
logarithmic derivative of the conductivity can be approximated by 
\begin{equation}
w(T) = c + d \cdot (T/T_{\rm m})^q\ ,
\end{equation}
where $T_{\rm m}$ is an experimentally accessible temperature, at which
$\sigma$ has been measured. 

Integration of Eq.\ (4) yields for  $q \ne 0$:
\begin{equation}
\ln \sigma(T_{\rm m}) - \ln \sigma(T) = c \cdot (\ln T_{\rm m} - \ln T)
+ d \cdot [1 - (T/T_{\rm m})^q] / q \ ,
\end{equation}
so that
\begin{equation}
\sigma(T) = \sigma(T_{\rm m}) \cdot (T / T_{\rm m})^c \cdot 
\exp \{-d [1 - (T/T_{\rm m})^q] / q \} \ . 
\end{equation}
For $q = 0$, we can assume $d = 0$, and get from Eq.\ (4):
\begin{equation}
\sigma(T) = \sigma(T_{\rm m}) \cdot (T / T_{\rm m})^c \ .
\end{equation}

Interpreting Eqs.\ (6,7), we have to discriminate between the eight 
situations listed in Table IV. According to this table, 
$w(T \rightarrow 0) = 0$ indicates metallic character of the 
conduction, whereas both diverging and finite positive 
$w(T \rightarrow 0)$ point out that the sample is an insulator, cf.\ 
Refs.\ \onlinecite{Moe.89,Hirs.etal,Thom.Holc,Rose.etal.94,Frie.etal}. 
In more detail, diverging and finite positive $w(T \rightarrow 0)$ are 
correlated with vanishing of $\sigma$ as $T \rightarrow 0$ according to
exponential and power laws, respectively, cf.\ Eqs.\ (6,7). Studying an
MIT, we can exclude the case that the limit $w(T \rightarrow 0) < 0$,
corresponding to $\sigma(T \rightarrow 0) = \infty$, as unphysical. 
However, this does of course not exclude negative $w$ at finite $T$, 
which can arise from weak antilocalization, or e.g.\ from 
electron-phonon scattering in good metals.

In this context, we may ask what the condition is for observing
exponential behavior, $\sigma(T) \propto \exp\{ -(T_0/T)^q \}$. The 
condition $T < T_0$ implies $w > q$. Thus there exists a low-$w$ range 
where, though increase of $w$ with decreasing $T$ would rule out 
metallic conduction, it would still not be possible to identify 
activated conduction by detecting exponential $\sigma(T)$. However,
assuming only monotonicity of $w$ as $T \rightarrow 0$, one reaches an
unambiguous decision.

The logarithmic derivative is also helpful in identifying the physical 
mechanisms involved \cite{Moe.etal.83,Moe.etal.85,Zabr.Zino} since it 
makes characteristic features of $\sigma(T)$ visible, which are not 
obvious from $\sigma(T)$ itself. An additional advantage of this 
quantity is that it is not influenced by geometrical errors, nor by the 
scaling of the resistivity mentioned above. 

Our method of data evaluation for analyses in terms of $w$ is based on 
that used in Ref.\ \onlinecite{Moe.88}: A linear fit of 
$\ln \sigma$ versus $\ln T$ for $k$ neighboring points is performed,
\begin{equation}
w_{\rm lf} =  \frac
{k \sum_i \ln \sigma_i \ln T_i - \sum_i \ln \sigma_i \sum_j \ln T_j}
{k \sum_i (\ln T_i)^2 - (\sum_i \ln T_i)^2} \,.
\end{equation}
In order to minimize the numerical error arising from the non-linearity
of the dependence of $\ln \sigma$ from $\ln T$, we relate this $w$ value 
to a mean temperature, $T_{\rm lf}$, defined by
\begin{equation}
\ln T_{\rm lf} = \frac
{k \sum_i (\ln T_i)^3 - \sum_i \ln T_i \sum_j (\ln T_j)^2}
{2 (k \sum_i (\ln T_i)^2 - (\sum_i \ln T_i)^2)}\,.
\end{equation}
Thus $w_{\rm lf}$ would exactly equal $w(T_{\rm lf})$ if $\ln \sigma$ 
were a second order polynomial of $\ln T$. Utilizing Eqs.\ (8,9) allows 
one to take into consideration rather broad $T$ intervals, so that the 
total error (numerical plus random) can be kept small.

The optimum value of $k$ depends on the accuracy and density of the data 
points, as well as on the structure of the measured curve.  We 
considered the following groups of neighboring data points: The 
measurements in the region 2 -- 300 K yielded data points with 
comparably high accuracy, but low density. From them, the logarithmic 
derivative was obtained using $k = 3$ below 7~K, and $k = 2$ above. In 
the $^3$He region, the density of data points is considerably higher,
thus 8 neighboring data points were considered in differentiating for 
all samples except for h, for which we used 4 points. The data obtained 
in the dilution refrigerator were evaluated using $k = 5$.

Fig.\ 7 gives a survey of the $T$ dependences of the logarithmic 
derivative.  Low-$T$ details become visible in Fig.\ 8. These 
figures show that both the sample sets A and B are internally 
consistent. However, there are also significant differences between 
these sets.

\subsection{Common features of the two series}

Consider first the common features: For both data sets, at sufficiently 
high $T$, all samples studied behave very similarly: $w(T)$ increases 
with increasing $T$. On the other hand, for the samples with the 
smallest Ni content, at low $T$, there is a pronounced increase of 
$w(T)$ with decreasing $T$, indicating activated conduction. The 
strength of this contribution decreases with increasing $x$. One 
common feature of all samples (with $w > 0$), which exhibit such an 
increase of $w$ with decreasing $T$, has to be stressed: this increase 
continues down to the lowest accessible $T$. The extrapolation / 
generalization of this finding will be a basic assumption of our data 
analysis.

The differing behavior at low and high $T$ must be caused by two 
different mechanisms being dominant at low and high $T$, respectively.
The low-$T$ mechanism related to increasing $w(T)$ as $T$ decreases is 
very likely a kind of hopping conduction, cf.\ Refs.\
\onlinecite{Moe.etal.83,Moe.etal.85,Rose.etal.97b,Zabr.Zino}.
Concerning the origin of the high-$T$ contribution we can only 
speculate. Such a high-$T$ contribution is common to many 
amorphous transition-metal semiconductor alloys \cite{Moe.85}.
Comparing energy scales, it was speculated that in 
a-Si$_{1-x}$Cr$_x$ this mechanism might be related to electron-phonon 
interaction \cite{Moe.etal.85}.

The feature, which we consider to be particularly interesting in Figs.\ 
7 and 8, is the minimum of $w(T)$ related to the crossover between these 
two mechanisms. This minimum is located at $T_{\rm min} = 150\ {\rm K}$ 
for sample a, which is the most insulating sample studied. The related 
minimum value of $w$ is $w_{\rm min} = 0.68$. The minimum shifts to 
lower $T$ with increasing Ni content. This behavior has to be expected 
since the characteristic hopping energy should tend to 0 as the 
transition is approached, so that hopping only becomes dominant at lower 
and lower $T$. 

Simultaneously with $T_{\rm min}$, $w_{\rm min}$ decreases: For the A 
samples 1 and 2, $T_{\rm min} = 6\ {\rm K}$ and 4~K, respectively, where 
$w_{\rm min} = 0.42$ and 0.32. For the B samples d and e,
$T_{\rm min} = 0.8\ {\rm K}$ and 0.2~K, respectively, where
$w_{\rm min} = 0.15$ and 0.06.

The four samples 1, 2, d, and e must be insulating because, for them, 
$w(T)$ increases with decreasing $T$ at the low-$T$ end of the $T$ range
considered. This increase cannot be understood in terms of metallic 
behavior according to Eq.\ (2). Moreover, because of the overall 
behavior of $w(T,x)$ (see first paragraph of this subsection), it is 
very likely that this trend continues as $T \rightarrow 0$ what 
indicates insulating character according to Tab.\ IV. Furthermore, even 
if $w(T)$ does not increase as $T \rightarrow 0$ but only remains 
constant, i.e., $w(T<T_{\rm min}) = w_{\rm min}$, the samples still have 
to be classified as insulating, because 
$\sigma(T<T_{\rm min}) \propto T^{w_{\rm min}}$, cf.\ previous 
subsection. This finding rules out the classification of these four 
samples as metallic according to the standard procedure used in Fig.\ 6.

However, for the samples 3, 4, and 5, $w(T)$ seems to tend to 0 as 
$T \rightarrow 0$, as well as for the samples g and h. (In this 
judgement, we presume that, since we study an MIT, the limit 
$w(T \rightarrow 0) \ge 0$, see previous subsection.) For sample f, we 
cannot decide whether or not $w(T \rightarrow 0) = 0$ although we 
investigated it down to 35~mK. Thus we conclude that the MIT should take 
place between samples 2 and 3 for set A, and between samples e and g for 
set B. 

In the calculation of the critical Ni contents, and of their possible 
errors, we take into account the Ni concentrations of the samples
bracketing the transition, the uncertainty of the $x$ values, and 
the influence of preparation conditions differing slightly from sample 
to sample in technology A. For sample sets A and B, we obtain 
$16.6 \pm 1\ {\rm at} \%$ according to our RBS scale, and 
$25 \pm 2\ {\rm at} \%$ according to the EDX scale of Ref.\ 
\onlinecite{Rose.etal.97a}, respectively. In order to compare both
values, we scale the EDX data by means of Table III: Presuming a linear 
relation between both EDX scales close to $x_{\rm c}$, and the RBS value 
to exceed the EDX value of this study by 5~at$\%$, one obtains 
$19 \pm 3\ {\rm at} \%$ as the critical Ni content of sample set B 
according to our RBS scale. Additionally performing the correction 
concerning the oxygen concentration, see Section III, we obtain similar
values for the critical concentration: $15.9 \pm 1$ and 
$17 \pm 3\ {\rm at} \%$ Ni for series A and B, respectively. Note 
however, that this agreement is not a necessity. On the contrary, since 
O is very reactive, it would not be surprising if the O contamination 
significantly influenced the gap states and shifted $x_{\rm c}$, cf.\ 
the role of H in a-Si based alloys \cite{Wrig.etal}.

The consideration of the samples bracketing the transition in both
sets shows that, at high $T$, samples with $x$ just below and above 
$x_{\rm c}$ exhibit very similar $T$ dependences of $w$ -- there 
is no experimental indication of a discontinuity in the $x$ dependence 
at finite $T$. Hence the respective substantial variations of $\sigma$ 
with $T$ on both sides of the MIT must originate from the same physical 
mechanism, which, crudely speaking, ``survives'' the transition. 
This high-$T$ mechanism yields a substantial contribution to $w$ 
down to the lowest $T$ studied, compare the samples closest to 
the MIT. Thus there is a superposition of the low- and high-$T$ 
contributions, whatever mechanisms they represent, in the whole $T$ 
interval investigated. This has an important consequence: In fitting 
simply to a pure hopping relation close to the MIT, or to Eq.\ (2), one 
might obtain a precise mathematical description, but the derived 
parameters would not have physical relevance. 

Concerning the high-$T$ mechanism, the comparison with 
a-Si$_{1-x}$Cr$_x$ given in Fig.\ 9 is interesting. For both the
a-Si$_{1-x}$Cr$_x$ sample $\delta$ and the a-Si$_{1-x}$Ni$_x$ sample 4, 
$w(300\ {\rm K}) \approx 0.5$. However, with decreasing $T$, the 
influence of the high-$T$ mechanism decreases considerably faster in 
sample $\delta$ than in sample 4: $w(30\ {\rm K}) = 0.04$ and 0.16 for 
samples  $\delta$ and 4, respectively. This faster decrease of the
influence of the high-$T$ mechanism is a general feature of 
a-Si$_{1-x}$Cr$_x$. It has two consequences: (i) Comparing insulating 
samples of both substances with the same $w_{\rm min}$, the minimum is 
reached in a-Si$_{1-x}$Cr$_x$ at clearly higher $T$ than in 
a-Si$_{1-x}$Ni$_x$, compare sample $\gamma$ with sample 2 in Figs.\ 9 
and 7a, respectively. (ii) In a-Si$_{1-x}$Cr$_x$, there are minima in 
$\sigma(T)$, i.e., zeros in $w(T)$, at significantly higher $T$ than in 
a-Si$_{1-x}$Ni$_x$, consider samples $\delta$ and $\varepsilon$, and see 
next subsection. Thus, in contrast to a-Si$_{1-x}$Ni$_x$, there is a 
broad, easily accessible low-$T$ region in a-Si$_{1-x}$Cr$_x$, where one 
can ignore this high-$T$ contribution, and study the pure low-$T$ 
mechanisms. 

\subsection{Differences between the two series}

Besides the common features described above, there are significant 
differences between the data sets A and B. These differences concern 
both the high-$T$ and the low-$T$ regions.

Comparing A and B samples with similar low-temper\-a\-ture $w(T)$ 
values, the A samples are found to exhibit larger high-temperature $w$ 
values, i.e., they are more strongly $T$ dependent. In detail, 
$w(300\ {\rm K},x_{\rm c}) = 0.67 \pm 0.06$ and $0.48 \pm 0.06$ for sets 
A and B, respectively. This might indicate a similarity to 
a-Si$_{1-x}$Cr$_x$, where the high-$T$ mechanism seems to lose its 
influence with increasing oxygen content of the films, compare Fig.\ 2 
of Ref.\ \onlinecite{Glad.etal} with Fig.\ 1 of Ref.\ 
\onlinecite{Moe.etal.85}. The same seems to be valid for 
a-Si$_{1-x}$Ni$_x$: The high-$T$ contribution has smaller 
influence for series B, for which the oxygen contamination is larger, as 
observed by EDX.

For samples exhibiting a decrease of $w(T)$ as $T \rightarrow 0$, 
this decrease is steeper for the A than for the B samples. Sample 3 even 
exhibits a zero of $w(T)$ at about 0.9~K, and for sample 4 a zero just 
below the lowest measuring $T$ is very likely. But such a zero was not 
found for the B samples within the $T$ region studied.  

Thus, in comparing the metallic samples of both sets, i.e., 3, 4, 5 and 
g, h, it is puzzling that for the A but not for the B samples there is 
visible a contribution to $\sigma(T)$ with negative $T$ derivative.
Under what conditions such a contribution exists is a question of 
current interest \cite{Albe.McLa}. E.g.\ such contributions were 
observed in a-Si$_{1-x}$Cr$_x$ \cite{Moe.etal.85}, as well as in 
heavily doped crystalline Si:P \cite{Rose.etal.81} and Ge:Sb 
\cite{Thom.etal.82}. It remains an open question whether the missing of 
this contribution in the B samples is real, or whether this contribution 
is obscured in set B by the low-$T$ part of the high-$T$ contribution. 
The latter hypothesis is suggested by the comparison in Fig.\ 7a which 
demonstrates that this low-$T$ part is considerably larger in the B 
samples than in the A samples.

Because our experimental data on the contribution with negative 
$\mbox{d} \sigma / \mbox{d} T$ are very sparse, we can only speculate on 
the physical mechanism behind. It is likely that its origin is the same 
as in a-Si$_{1-x}$Cr$_x$ \cite{Moe.etal.85,Moe.90a}, see also Fig.\ 
9, heavily doped crystalline Si:P \cite{Rose.etal.81}, and Ge:Sb 
\cite{Thom.etal.82}, that means probably some interplay of 
electron-electron interaction and disorder. However, the interpretation 
in terms of precursors of a metal-superconductor transition as in 
a-Si$_{1-x}$Au$_x$, where both the pure constituents do not become 
superconducting \cite{Nish.etal}, cannot be ruled out yet. Further 
investigations at lower $T$ are desirable.

Since great care was taken for both sample sets concerning homogeneity,
we ascribe the differences stressed to the influence of differing 
preparation conditions. Significant contamination by crucible material 
could be excluded for both sample sets: In a related GDOES study, the 
contamination of an A sample by copper from the crucible was found to be 
below 500 ppm. Concerning a crucible-material-contamination check for 
the B samples see Ref.\ \onlinecite {Rose.etal.97a}. Thus only three 
possibilities seem to remain:\\
-- Contamination by any constituents from the residual gas,\\
-- differing surface diffusion during deposition arising from the 
   differences in rate and / or temperature, and\\
-- concentration fluctuations on a mesoscopic length scale arising in 
   the technology B, cf.\ Section III.\\
Further investigations are needed to clarify this point.

\section{Discussion: The qualitative character of the MIT}

\subsection{Missing plateau in $w(T)$}

The common features of both sample sets suggest an important qualitative
conclusion concerning the character of the MIT, as will be explained in 
this section. Figs.\ 7 and 8, together with Tables I and II, show that 
$w(T = {\rm const.},x)$ decreases monotonically with increasing $x$, at 
least as long as $w > 0$. Thus there must be a separatrix (separating 
line), above which insulating and below which metallic samples are 
located in these graphs. More accurately, instead of a separatrix, there 
could also exist a separating strip, but this would imply a 
discontinuity in $\sigma(T = {\rm const.},x)$ for which there is no 
experimental evidence in this experiment, nor in the literature.

Now we compare the consequences of the common assumptions with 
experiment. For that we assume that the transition is continuous, as
described by Eq.\ (1), and that, sufficiently close to the transition, 
below some temperature $T^\star$, Eq.\ (2) holds. These equations imply 
three conclusions on $w(T,x)$:

(i) $w(T,x_{\rm c}) = p$ since $a(x_{\rm c}) = 0$. Thus the separatrix 
should be parallel to the $T$ axis below $T^\star$. 

(ii) For metallic samples close to the MIT, $w \approx p$ as long as 
$a(x) \ll b(x) \cdot T^p$. In other words, since $a(x)$ vanishes as 
$x \rightarrow x_{\rm c}$, the logarithmic derivative 
$w(T,x = {\rm const.})$ should exhibit a plateau below $T^\star$, which 
extends to lower $T$ the closer $x$ to $x_{\rm c}$. 

(iii) A plateau should also be present on the insulating side close to 
$x_{\rm c}$ below $T^\star$ for the following reasons. Since the 
charateristic hopping energy very likely tends continuously to 0 as 
$x \rightarrow x_{\rm c}$, the crossover temperature, below which the 
influence of hopping dominates the $T$ dependence of $w$, vanishes too.
Continuity of $\sigma(T = {\rm const.},x)$ leads to the expectation that 
$w \approx {\rm const.}$ above that crossover temperature.

However, according to our experimental data, such a plateau is missing.
Instead, for arbitrary fixed $T$, $w(T,x_{\rm c})$ must be smaller than 
$w(T)$ of samples 2 and e, but exceed $w(T)$ of samples 3 and g for 
sample sets A and B, respectively. Thus, $w(T,x_{\rm c})$ decreases with 
decreasing $T$, at least down to 0.2 K, where $w(T,x_{\rm c}) < 0.06$. 
There is no experimental hint for $w(T,x_{\rm c})$ at even lower $T$ to 
rise again up to one of the theory values 1/2 or 1/3, and to saturate 
there; for $x$ close to $x_{\rm c}$, such a behavior would imply the 
existence of metallic $w(T)$ curves with a low-$T$ maximum, so that it 
would contradict the assumption on monotonicity of $w(T)$ as $T$ 
vanishes for samples with $w > 0$ and $\mbox{d} w / \mbox{d} T < 0$, 
concluded in Section V.C from our experimental data. Hence, at least one 
of the standard descriptions Eqs.\ (1,2) is very likely not valid: 
Either the relevant mechanism at the transition is related to a value of 
$p$ which is far smaller than expected, or the transition is not 
continuous at $T = 0$, since $w(T \rightarrow 0,x_{\rm c}) = 0$ 
indicates finite $\sigma(T \rightarrow 0,x_{\rm c}+0)$ according to 
Table IV.

\subsection{Two simple models}

To illustrate the problem from a different perspective, we study 
$w(T,x)$ for two qualitative empirical models. They are constructed so 
that continuity at $x_{\rm c}$ for arbitrary finite $T$ and monotonic 
behavior of $\sigma(T = {\rm const.},x)$ are guaranteed. 

First we assume the transition to be continuous also at $T = 0$. One of 
the simplest suitable models is 
\begin{equation}
\sigma(T,x) = \left\{ \begin{array}{l} 
T^{1/2} \exp\{-(T_0(x)/T)^{1/2}\}\\ a(x) + T^{1/2}
\end{array} \quad\quad\mbox{for}\quad\quad 
\begin{array}{l} x < x_{\rm c} \\ x \ge x_{\rm c} \end{array} \,, 
\right. 
\end{equation}
where $T_0(x \rightarrow x_{\rm c} - 0) = 0$, and 
$a(x \rightarrow x_{\rm c} + 0) = 0$. For simplicity, all quantities 
are given in dimensionless form. 

Corresponding $w(T)$ curves are presented in Fig.\ 10. Note that there 
are no pieces with $w < p = 1/2$ which belong to insulating samples, and 
that, if $w < p$, $\mbox{d} w / \mbox{d} T$ is always positive. 

Next we assume the transition to be discontinuous, but only at $T = 0$. 
We consider again a very simple qualitative model:
\begin{equation}
\sigma(T,x) = \left\{ \begin{array}{l} 
(1 + T^{1/2}) \exp\{-(T_0(x)/T)^{1/2}\}\\ a(x) + T^{1/2} 
\end{array} \quad\quad\mbox{for}\quad\quad 
\begin{array}{l} x < x_{\rm c} \\ x \ge x_{\rm c} \end{array} \,,\right.
\end{equation}
where $T_0(x \rightarrow x_{\rm c} - 0) = 0$, and
$a(x \rightarrow x_{\rm c} + 0) = 1$. In order to illustrate how this
model unifies a discontinuity of $\sigma(T={\rm const.},x)$ at $T = 0$ 
with continuity for $T > 0$, we present $\sigma(T,x={\rm const.})$ and 
$\sigma(T={\rm const.},x)$ in Figs.\ 11a and 11b, respectively. The 
latter graph is based on the following additional assumptions: 
$x_{\rm c} = 0.5$, $T_0(x) = 30 \cdot (x_{\rm c}-x)^2$, and 
$a(x) = 1+3 \cdot (x-x_{\rm c})$, where using 2 as critical exponent of 
$T_0$ is motivated by \onlinecite{Zabr.Zino}; however, the qualitative 
properties of the model considered do not depend on the numbers chosen. 
Figs.\ 11a and 11b demonstrate two remarkable features of the model 
(11): (i) If some low-$T$ range, e.g.\ the shaded area, is masked, a 
misinterpretation of the activated curves closest to the MIT as metallic 
ones is tempting, and one could even conclude that the transition is 
continuous. (ii) The singularity of $\sigma(0,x)$ at $x_{\rm c}$ evolves 
with decreasing $T$ because $\sigma(T={\rm const.},x)$ becomes 
increasingly steep just below $x_{\rm c}$.

Fig.\ 12 shows the $w(T)$ curves for model (11). Contrary to Fig.\ 10, 
this figure includes minima of $w(T,x = {\rm const.})$ with very small 
values of $w_{\rm min}(x)$ at $T_{\rm min}(x)$, where $T_{\rm min}(x)$ 
as well as $w_{\rm min}(x)$ vanish as $x \rightarrow x_{\rm c}$. The 
specific point is that $\sigma(0,x_{\rm c}-0)$ equals 
$\sigma(0,x_{\rm c})$ for model (10), but not for (11). In both cases, 
$w(0,x_{\rm c}-0) = \infty$, and $w(0,x_{\rm c}+0) = 0$;  the limit 
$w(T \rightarrow 0,x_{\rm c})$ equals 1/2 for model (10) and 0 for 
model (11).

Concerning the existence of minima in $w(T,x = {\rm const.})$, Fig.\ 12 
resembles far more the experimental figures than Fig.\ 10. Thus, a 
discontinuity of $\sigma(0,x)$ is suggested, provided $w(T,x_{\rm c})$ 
does not finally saturate (at some value far lower than theoretically 
expected) below the lowest experimentally accessible $T$ as $T$ 
decreases. However, this result was obtained using a very simple model. 
The question of its general validity is considered in the next 
subsection.

\subsection{Mathematical view}

The argumentation leading from the discontinuous character of the MIT, 
as modeled by Eq.\ (11), to the qualitative character of $w(T,x)$ shown 
in Fig.\ 12 can easily be reversed: Figs.\ 7 and 8 suggest the 
assumption that $w(T,x_{\rm c}) \rightarrow 0$ as $T \rightarrow 0$. 
Moreover, we only need the experimentally demonstrated monotonicity of 
$\sigma(T={\rm const.},x)$, that is $\mbox{d} \sigma / \mbox{d} x > 0$. 
We suppose it to be valid down to arbitrarily low $T$. 

Formulated in mathematical language, Figs.\ 7 and 8, together with Table 
IV, suggest that the samples can be discriminated according to the 
behavior of $w(T,x)$ for vanishing $T$ to belong to either of two groups 
with $x < x_{\rm c}$ and $x \ge x_{\rm c}$, respectively:\\ 
(i) For samples with ``activated'' conduction, classified as 
``insulating'', there is a positive $w_1(x)$, such that 
\begin{equation}
w(T,x) \ge w_1(x)\ \ {\rm for\ all}\ \ T < T_{\rm m}\,,
\end{equation}
where $T_{\rm m}$ is some experimentally accessible temperature.\\
(ii) For ``metallic'' samples, there are numbers $w_0$ and $q$ 
(independent of $x$) such that
\begin{equation}
w(T,x) \le w_0 \cdot (T/T_{\rm m})^q\ \ {\rm for\ all}\ \ 
T < T_{\rm m}\,.
\end{equation}

We assume that we have determined $\sigma(T_{\rm m},x)$ experimentally.
For the ``insulating'' $x$ region, integration of Eq.\ (12) leads to 
\begin{equation}
\sigma(T,x) \le \sigma(T_{\rm m},x) \cdot (T/T_{\rm m})^{w_1(x)}\,,
\end{equation}
so that 
\begin{equation}
\sigma(0,x) = 0\ \ {\rm for\ all}\ \ x < x_{\rm c}\,.
\end{equation} 
Now we turn to the ``metallic'' region. For $x = x_{\rm c}$, integration 
of Eq.\ (13) yields
\begin{equation}
\ln \sigma(T_{\rm m},x_{\rm c})-\ln \sigma(T,x_{\rm c}) \le 
w_0 \cdot [1 - (T/T_{\rm m})^q]/q\,,
\end{equation}
so that 
\begin{equation}
\sigma(0,x_{\rm c}) \ge 
\sigma(T_{\rm m},x_{\rm c}) \cdot \exp(-w_0/q) > 0\,.
\end{equation}
That means $\sigma(0,x_{\rm c})$ must be finite. (In our case, Fig.\ 8 
leads to the estimates $w_0 = 0.05$ and $q = 1/2$ for 
$T_{\rm m} = 1\ {\rm K}$, so that $\exp\{-w_0/q\} \approx 0.9$.) 
Finally, due to the monotonicity of $\sigma(T={\rm const.},x)$, we get 
\begin{equation}
\sigma(0,x) > \sigma(T_{\rm m},x_{\rm c}) \cdot \exp(-w_0/q)\ \ 
{\rm for\ all}\ \ x > x_{\rm c}\,.
\end{equation}

Hence, the classifications of the samples according to 
$w(T \rightarrow 0,x)$ and $\sigma(T \rightarrow 0,x)$, respectively,
are identical, and we arrive at the following conclusion: Provided 
our qualitative assumptions on the limiting behavior of $w(T,x_{\rm c})$ 
and the monotonicity of $\sigma(T = {\rm const.},x)$ are justified, 
there must be a finite minimum metallic conductivity.

\subsection{Consideration of counterarguments}

We must, however, look back at our arguments to see what would be
necessary for the transition nevertheless to be continuous, in which 
manner condition (13) would have to be weakened. Two possibilities have 
already been mentioned: (i) The behavior of $w(T,x)$ might qualitatively 
change at temperatures below the lowest one accessible to us. (ii) At 
$x_{\rm c}$, the parameter $c$ of Eq.\ (4) might have a small but finite 
value, far smaller than the values 1/2 and 1/3, which were predicted 
theoretically in Refs.\ \onlinecite{Alts.Aron,News.Pepp} and used in 
many experimental studies in the literature. 

A third possibility is found when referring to Eq.\ (5), which, for 
$c = 0$, simplifies to
$\ln \sigma (T=0) = \ln \sigma (T_{\rm m}) - d/q$. For no discontinuity 
to occur, it has to be possible to reach any (arbitrarily small) 
$\sigma (0,x_{\rm c}+0)$. The first term, $\ln \sigma (T_{\rm m})$, is 
finite due to the monotonicity of $\sigma(T_{\rm m},x)$ and the finite 
$\sigma$ values observed in the insulating region at $T_{\rm m}$. Thus 
for no discontinuity to occur, $d/q$ has to be infinite. Since 
$d = w(T_{\rm m})$, the exponent $q$ must be infinitely small. For $T$ 
independent $q$, this situation is equivalent to the case (ii) above, 
that is the situation of a small but finite $c$. The only remaining 
possibility is that of a $T$ dependent $q$, vanishing as 
$T \rightarrow 0$ for $x = x_{\rm c}$. This would imply a non-power law 
behavior of $\sigma(T,x_{\rm c})$. If one would nevertheless try to fit 
$\sigma \propto T^p$, the effective exponent $p$ would be $T$ dependent,
and approach 0 as $T \rightarrow 0$.

However, we know of no physical argument that suggests that the power 
law (which is derived by perturbation arguments) should weaken at low 
$T$, and we did not obtain any data that suggest that there is a 
tendency towards such a change at lowest accessible $T$. Therefore, for 
$T=0$, although continuity across the transition is theoretically 
possible, experimental evidence appears to indicate that the transition 
is, in fact, discontinuous.

Finally, the question remains whether or not this result might be an 
artifact arising from any sample inhomogeneities. We utilized all 
available experimental possibilities to control and to minimize 
macroscopic inhomogeneities. Nevertheless, surely some small composition 
gradients remain. Calculating the $T = 0$ conductance of samples with 
constant $x$ gradients perpendicular and parallel to the current 
direction, respectively, assuming Eq.\ (1) to be valid, one obtains the 
following results: In the former case, the transition is ``smeared 
out'', but in the latter case, provided the exponent $\nu$ in Eq.\ (1) 
satisfies $0 < \nu < 1$, a sharp transition would result. However, 
according to the preparation technologies used, the inhomogeneities 
should be considerably smaller parallel to the current direction than 
perpendicular to it. Thus the influence of a parallel $x$ gradient is 
probably overcompensated by the perpendicular inhomogeneities. Moreover, 
the hypothesis that the discontinuous transition for $T = 0$ might be 
only an artifact originating from a parallel gradient, is ruled out by 
the behavior of $w(T,x = {\rm const.})$ for the region where 
$a(x) \ll b(x) \cdot T^p$. The logarithmic derivative would have to 
exhibit a plateau $w(T,x_{\rm c}) \approx p$ here according to the 
gradient hypothesis, but we did not find one.

\subsection{Minimum metallic conductivity}

In summary, it is very likely that $\sigma(0,x)$ exhibits a 
discontinuity in a-Si$_{1-x}$Ni$_x$, corresponding to Mott's idea of a 
minimum metallic conductivity $\sigma_{\rm mm}$ \cite{Mott.72}, but 
$\sigma(T = {\rm const.} > 0,x)$ should be continuous for arbitrary 
finite $T$. However, we cannot exclude the possibility that 
$w(T,x_{\rm c})$ changes its behavior, in particular that it 
saturates at some finite value, at temperatures lower than those
experimentally accessible to us.  This would correspond to the $T$ 
exponent in Eq.\ (2) becoming much smaller than expected theoretically. 
Thus a conclusion with complete certainty is impossible. 

Our experimental data only allow us to determine bounds for 
$\sigma_{\rm mm}$. From the sample sets A and B we get 
$20\ \Omega^{-1}{\rm cm}^{-1} < \sigma_{\rm mm} < 
65\ \Omega^{-1}{\rm cm}^{-1}$, and
$25\ \Omega^{-1}{\rm cm}^{-1} < \sigma_{\rm mm} < 
70\ \Omega^{-1}{\rm cm}^{-1}$, respectively. It is instructive to 
compare these data with Mott's theoretical result for $\sigma_{\rm mm}$
\cite{Mott.72,Mott.Davi.79}, though electron-electron interaction and 
weak localization effects are neglected in its derivation. According to 
Ref.\ \onlinecite{Mott.Davi.79}, 
$\sigma_{\rm mm} = 0.026\, e^2/(\hbar a)$, where $a$ denotes the 
distance between neighboring impurity atoms, i.e., Ni atoms, at the MIT. 
Taking the atomic density of the samples to be similar that in 
crystalline Si, we get $a \approx 5\ {\rm \AA}$. This estimate fits with 
the Mott-Edwards-Sienko criterion, 
$a_{\rm H}^\ast \cdot n_{\rm c}^{1/3} \approx 0.25$ with 
$a_{\rm H}^\ast$ being the Bohr orbit radius of an isolated center and
$n_{\rm c}$ the critical impurity density 
\cite{Mott.56,Edwa.Sien.78,Edwa.Sien.83}: Approximating $a_{\rm H}^\ast$ 
by half the nearest neighbor distance of crystalline Ni, 
$a_{\rm H}^\ast \approx 1.24\ {\rm \AA}$, yields just the expected 
product value.  Finally, we obtain 
$\sigma_{\rm mm} \approx 120\ \Omega^{-1}{\rm cm}^{-1}$. This result 
exceeds  our lower and upper bounds on $\sigma_{\rm mm}$ by factors of 
roughly 5 and 2, respectively. Note however that the above consideration
does not take account of the possible d character of the Ni states;
moreover, the dimensionless parameters have some uncertainty.

For comparison, for a-Si$_{1-x}$Cr$_x$, the conductivity studies lead to
$\sigma_{\rm mm} \approx 250\ \Omega^{-1}{\rm cm}^{-1}$ 
\cite{Moe.etal.83,Moe.etal.85}. In that substance, annealing causes 
a conductivity decrease \cite{Moe.etal.85}, but the question whether 
or not annealing also causes a decrease of $\sigma_{\rm mm}$ is still 
open. However, due to the differences between the phase diagrams of 
the systems Si-Ni and Si-Cr, see Section I, the following hypothesis is 
suggested: If, instead of the Ni-Ni distance, the distance between 
metallic NiSi$_2$ grains were the relevant length for $\sigma_{\rm mm}$, 
the relatively small value of $\sigma_{\rm mm}$ in a-Si$_{1-x}$Ni$_x$ 
might result from the formation of NiSi$_2$ crystalline regions with a 
diameter of the order of 1 to 2~nm, cf.\ 
\onlinecite{Edwa.etal,Raap.etal}. Corresponding structural studies would 
be interesting.

\section{Conclusions}

The main result of our study is the following: the $T$ dependence of 
the logarithmic temperature derivative of the conductivity, 
$w(T,x) = \mbox{d} \ln \sigma / \mbox{d} \ln T$, for 
a-Si$_{1-x}$Ni$_x$ qualitatively differs from the predictions of
commonly accepted theory. According to this theory, $\sigma(0,x)$ is 
continuous, and, on the metallic side of the MIT, 
$\sigma(T,x) = a(x) + b(x) \cdot T^p$, where $p$ should equal 1/2 or 
1/3. Thus $w(T,x_{\rm c}) = p$, and, for $w < p$, only positive 
$\mbox{d} w / \mbox{d} T$ is expected. However, we observed 
characteristic minima of $w(T,x = {\rm const.})$, where not 
only $T_{\rm min}$ but also $w_{\rm min}$ seem to tend to 0 as 
$x \rightarrow x_{\rm c}$ from the insulating side. Below $T_{\rm min}$, 
$w$ was observed to increase monotonically with decreasing $T$. The 
detailed analysis of this feature favors the discontinuous character of 
the MIT at $T = 0$, in the sense of Mott's original prediction of a 
finite minimum metallic conductivity.

The question arises why a minimum metallic conductivity was found in 
a-Si$_{1-x}$Ni$_x$, and previously in a-Si$_{1-x}$Cr$_x$
\cite{Moe.etal.83,Moe.etal.85,Moe.87,Moe.90a,Moe.90b}, 
but not in many of the other substances, e.g.\ crystalline Si:P, 
investigated so far. Of course, at the present stage, one cannot exclude 
the possibility of a-Si$_{1-x}$Ni$_x$ and Si:P belonging to different 
universality classes. Simultaneously, however, the question arises
as to whether an analysis using $w(T,x)$ of previous experiments might 
reveal similar contradictions between standard description and 
experimental data as we found here. But, for the lack of sufficiently 
detailed information on these investigations, this question has to be 
postponed to future studies. Such analyses should be highly interesting 
since  samples with $w < 1/3$ and negative $\mbox{d} w / \mbox{d} T$ at 
the lowest $T$ were also found in experiments on crystalline Si:As 
\cite{Shaf.etal} and Si:(P,B) \cite{Thom.Holc}, on amorphous 
Si$_{1-x}$Cr$_x$ \cite{SiCrscale} and Si$_{1-x}$Mn$_x$ \cite{Yaki.etal}, 
as well as on granular Al-Ge \cite{Rose.etal.94}. But there is also one 
case in the literature, where, for such a sample, $w$ first increases 
and then clearly decreases as $T$ vanishes: the a-Cr-SiO$_x$ sample 7 in 
Ref.\ \onlinecite{Vinz.etal} (see Fig.\ 2 of that work). Moreover, to 
the best of our knowledge, none of the experiments, which are 
interpreted as proof of the continuity of the MIT at $T = 0$, excludes 
that the ``metallic'' samples very close to the MIT could be insulating 
with a characteristic hopping temperature smaller than the lowest 
experimentally accessible $T$ \cite{Cast,Moe.89}.

Additionally to the result concerning the character of the MIT, we 
consider the following points as important conclusions:\\
-- Although both the sample sets A and B, prepared with different 
technologies, differ significantly concerning $\sigma(T,x)$, their 
critical Ni concentrations are close: $15.9 \pm 1$ and 
$17 \pm 3\ {\rm at} \%$, respectively, related to our RBS scale.\\ 
-- A high-$T$ mechanism, which ``survives'' the transition, masks the 
specific localization features to a large extent in both sample sets. It 
remains relevant down to temperatures of several hundred mK. Thus 
determining theoretical parameters by fitting conductivity formulae 
which account only for one mechanism is not justified for 
a-Si$_{1-x}$Ni$_x$ within the considered $(T,x)$ region.\\
-- For a-Si$_{1-x}$Ni$_x$ also, it is possible to prepare metallic 
samples with the conductivity increasing with decreasing $T$. However,
the influence of the mechanism from which this feature originates is 
weak, and it is directly visible only for certain preparation 
conditions.

We would also like to stress a technological consequence of our STM
investigations: When evaporating from two crucibles, the surface 
roughness of the films is large enough to lead to substantial 
concentration fluctuations on a mesoscopic length scale, caused by 
fluctuating incidence angles. To what extent these fluctuations are 
washed out by diffusion is an open question.

Finally, we turn to the question of what can be learned concerning 
the nature of the MIT in a-Si$_{1-x}$Ni$_x$ from the phenomenological 
results obtained above. Our findings are in disagreement with 
conclusions drawn from the treatment of the MIT in terms of 
one-parameter scaling theory \cite{Abra.etal} of Anderson localization 
of non-interacting electrons. But, on their own, they cannot explain 
why this description fails. For example, they do not tell us whether it 
is because electron-electron interaction had to be taken into account, 
nor whether two (or even more) scaling parameters would be needed to get 
an adequate result. Therefore, measurements of other quantities have to 
be consulted. XPS, UPS, and XES studies \cite{Isob.etal,Gheo.etal}, as 
well as optical measurements \cite{Davi.etal,Asal.etal} have shown how 
the gap in the density of states shrinks and finally closes with 
increasing Ni content, and how the Fermi energy is shifted in this 
process. However, for reasons of resolution, such experiments do not 
yield information on the important immediate vicinity of the MIT since 
the characteristic energy (temperature) of the hopping conduction 
vanishes as $x$ approaches $x_{\rm c}$ from below. 

A helpful hint comes from low-$T$ hopping conduction deep in the 
insulating region \cite{Abke.etal.92a,Abke.etal.92b} -- there, it has a 
far stronger influence on $w$ than the high-temperature mechanism which
survives the MIT. Since the value of the hopping exponent $q$ in the
approximation $\sigma \propto \exp\{-(T_0/T)^q\}$ is close to 1/2, and 
thus considerably exceeds Mott's result 1/4 for non-interacting 
electrons \cite{Mott.68}, electron-electron interaction probably is 
important -- the question whether or not this value can be interpreted 
in terms of hopping in the Coulomb gap is still under controversial 
debate, see e.g.\ \onlinecite{Efro.Shkl,Adki,Poll.98}. Thus, it seems
likely that the MIT arises from interwoven localization and 
electron-electron interaction processes, cf.\ the speculations in 
Ref.\ \onlinecite{Moe.85}. A theoretical approach in this direction 
was very recently published by Chitra and Kotliar \cite{Chit.Kotl}. 
These authors incorporate the {\it long-range} Coulomb interaction into 
dynamical mean-field theory, and obtain the result that the MIT should 
be discontinuous in two- and three-dimensional systems. However, if this 
hypothesis is true, we would be left with a new puzzle: Why does Mott's 
estimate of the minimum metallic conductivity, the derivation 
\cite{Mott.Davi.79} of which neglects electron-electron interaction, 
still yield a reasonable value for $\sigma_{\rm mm}$ in
a-Si$_{1-x}$Ni$_x$?

\section*{Acknowledgments}

This work was supported by the German-Israeli Foundation for Scientific
Research and Development. We are indebted to many colleagues for 
helpful discussions, in particular to B.~Altshuler, W.~Br\"uckner, 
G.~Diener, A.~Finkel'stein, P.~H\"aussler, B.~Kramer, C.~Lauinger, 
Z.~Ovadyahu, G.~Reiss, and U.~R\"ossler. We are much obliged to 
A.~Isobe, M.~Yamada, and K.~Tanaka for sending us their $\sigma(T)$ 
data of sputtered films. Special thanks go to S.~Geoghegan for the 
support in presenting three-dimensional STM data by means of 
Mathematica.

\begin{table}
\begin{tabular}{ccccc}
Sample&$x/{\rm at} \%\ {\rm Ni}$&$t/{\rm nm}$&
$\sigma(300\ {\rm K})/\Omega^{-1}{\rm cm}^{-1}$&
$\sigma(4.2\ {\rm K})/\Omega^{-1}{\rm cm}^{-1}$\\
\hline
1&14.9&159&150&15\\
2&16.4&183&160&20\\
3&16.7&179&245&64\\
4&18.9&181&340&145\\
5&22.1&162&585&330\\
6&55.9&175&--&--\\
\end{tabular}
\caption{
Ni content $x$, film thickness $t$, conductivity $\sigma$ at room
and liquid-helium temperatures for sample set A. The $x$ values are 
obtained from the Ni:Si ratio measured by RBS within this work; their 
uncertainty amounts to 0.3~at$\%$ for samples 1 -- 5, and to 0.5~at$\%$ 
for sample 6.
}
\end{table}

\begin{table}
\begin{tabular}{ccccc}
Sample&$x/{\rm at} \%\ {\rm Ni}$&$t/{\rm nm}$&
$\sigma(300\ {\rm K})/\Omega^{-1}{\rm cm}^{-1}$&
$\sigma(4.2\ {\rm K})/\Omega^{-1}{\rm cm}^{-1}$\\
\hline
a (23)&19.5&121&42&1.0\\
b (22)&20.3&122&54&2.7\\
c (21)&21.2&124&73&8.0\\
d (20)&22.3&124&88&18\\
e (19)&23.5&123&120&30\\
f (18)&24.8&122&160&60\\
g (17)&26.4&120&190&82\\
h (16)&28.2&118&280&160\\
i (8)&42.9&89&790&--\\
j (--)&25.3&--&--&--\\
\end{tabular}
\caption{
Ni content $x$, film thickness $t$, conductivity $\sigma$ at room
and liquid-helium temperatures for sample set B. The sample names in 
parentheses are the notations used in Ref.\ 
{\protect \onlinecite{Rose.etal.97a,Rose.etal.97b}}. Sample j, 
positioned between samples f and g during film deposition, was 
prepared specifically for EDX and RBS analyses within that work. The $x$ 
values are the EDX data published in either Tab.\ I of those papers.
}
\end{table}

\begin{table}
\begin{tabular}{ccccc}
Sample&RBS-R&RBS-F&EDX-D&EDX-W\\
\hline
3&16.7&--&11.6&--\\
3&16.7&--&12.0&--\\
6&55.9&--&51.7&--\\
f&--&--&14.0&24.8\\
h&--&--&20.3&28.2\\
i&--&--&37.8&42.9\\
j&--&20.0&--&25.3\\
\end{tabular}
\caption{
Comparison of Ni contents as determined by RBS in Rossendorf (RBS-R)
and Faure (RBS-F), and by EDX in Dresden (EDX-D) and Witwatersrand
(EDX-W). The RBS-R and EDX-D data were obtained within this work,
whereas the RBS-F and EDX-W results are taken from Ref.\
{\protect \onlinecite{Rose.etal.97a}}, -- unfortunately, Ref.\
{\protect \onlinecite{Rose.etal.97a}} contained a mistake in 
evaluating $x$ from the Si:Ni ratio obtained by RBS-F, which is 
corrected here. All Ni contents were obtained from the Si:Ni ratios, 
ignoring impurity contamination. The random error of the EDX measurement 
within this study does not exceed 0.5~at$\%$ Ni. The first two lines 
serve also as reproducibility check of EDX-D; these measurements were 
performed fully independently at two different days, when different 
pieces of the film were investigated. 
}
\end{table}

\begin{table}
\begin{tabular}{cccccc}
$q$&$c$&$d$&$w(T \rightarrow 0)$&$\sigma(T \rightarrow 0)$&character\\
\hline
$< 0$&arbitrary&$< 0$&$-\infty$&$\infty$&ideal metal\\ 
$< 0$&arbitrary&$> 0$&$\infty$&0&insulator\\ 
$= 0$&$< 0$&$= 0$&$< 0$&$\infty$&ideal metal\\
$= 0$&$= 0$&$= 0$&$= 0$&finite&real metal\\
$= 0$&$> 0$&$= 0$&$> 0$&0&insulator\\
$> 0$&$< 0$&arbitrary&$< 0$&$\infty$&ideal metal\\ 
$> 0$&$= 0$&arbitrary&$= 0$&finite&real metal\\
$> 0$&$> 0$&arbitrary&$> 0$&0&insulator\\ 
\end{tabular}
\caption{
Behavior of $\sigma(T \rightarrow 0)$, and character of the conduction
process in dependence on the parameters in Eq.\ (4), modelling 
$w(T \rightarrow 0)$. Here, ``ideal metal'' stands for 
$\sigma(T \rightarrow 0) = \infty$ (no impurity scattering). For 
simplicity, if $q = 0$, we assume $d = 0$; moreover, we consider $d = 0$ 
only for $q = 0$.
}
\end{table}

\begin{figure}
\caption{
Experimental parameter plane: The insulating region, where only 
activated conduction occurs, is marked by I, the metallic region by M, 
and the lowest experimentally accessible temperature by $T_{\rm lea}$. 
Measuring $\sigma(T,x = {\rm const.})$ means to obtain data points 
($\bullet$, $\times$) along vertical lines in this graph. The 
characteristic hopping temperature, $T_{\rm hop}(x)$, is marked by a 
full line. Only for $T < T_{\rm hop}$, $\sigma$ depends exponentially on 
$T$. For $T > T_{\rm hop}$, comparably flat, non-exponential $\sigma(T)$ 
dependences are expected. Thus, for $\bullet$, depending on $T$, both 
exponential and non-exponential behavior are observed. However, for 
$\times$, only non-exponential $\sigma(T,x = {\rm const.})$ are found, 
although this sample belongs to the insulating region, too. The latter 
problem occurs in the whole interval $[x^\star,x_{\rm c})$.
}
\label{fig1}
\end{figure}

\begin{figure}
\caption{
Glow discharge optical emission spectroscopy (GDOES) depth profiles for 
the elements B, C, Ni, and Si, i.e., intensity $I$ versus sputtering 
time $t$, for two samples prepared by technologies A and B, 
respectively. (a) presents the analysis of sample 5, whereas (b) shows 
data for a sample prepared specifically for this analysis together with 
samples a -- i; due to the simultaneous preparation, (b) is 
representative for the whole sample set B. We made use of the following 
lines: B - 208.9~nm, C - 156.1~nm, Ni - 349.2~nm, and Si - 288.1~nm. The 
intensity scales were adjusted to reach a high resolution, thus the 
units differ from element to element. In (a), the sudden increase of the 
B signal indicates the moment when the bottom of the sputtering crater 
reaches the glass substrate. 
}
\label{fig2}
\end{figure}

\begin{figure}
\caption{
STM image of the surface topography of sample 3, measured with a current 
of 0.5~nA and a bias of 8~V. The figure was obtained using a slope 
correction and a slight smoothing of the row data.
}
\label{fig3}
\end{figure}

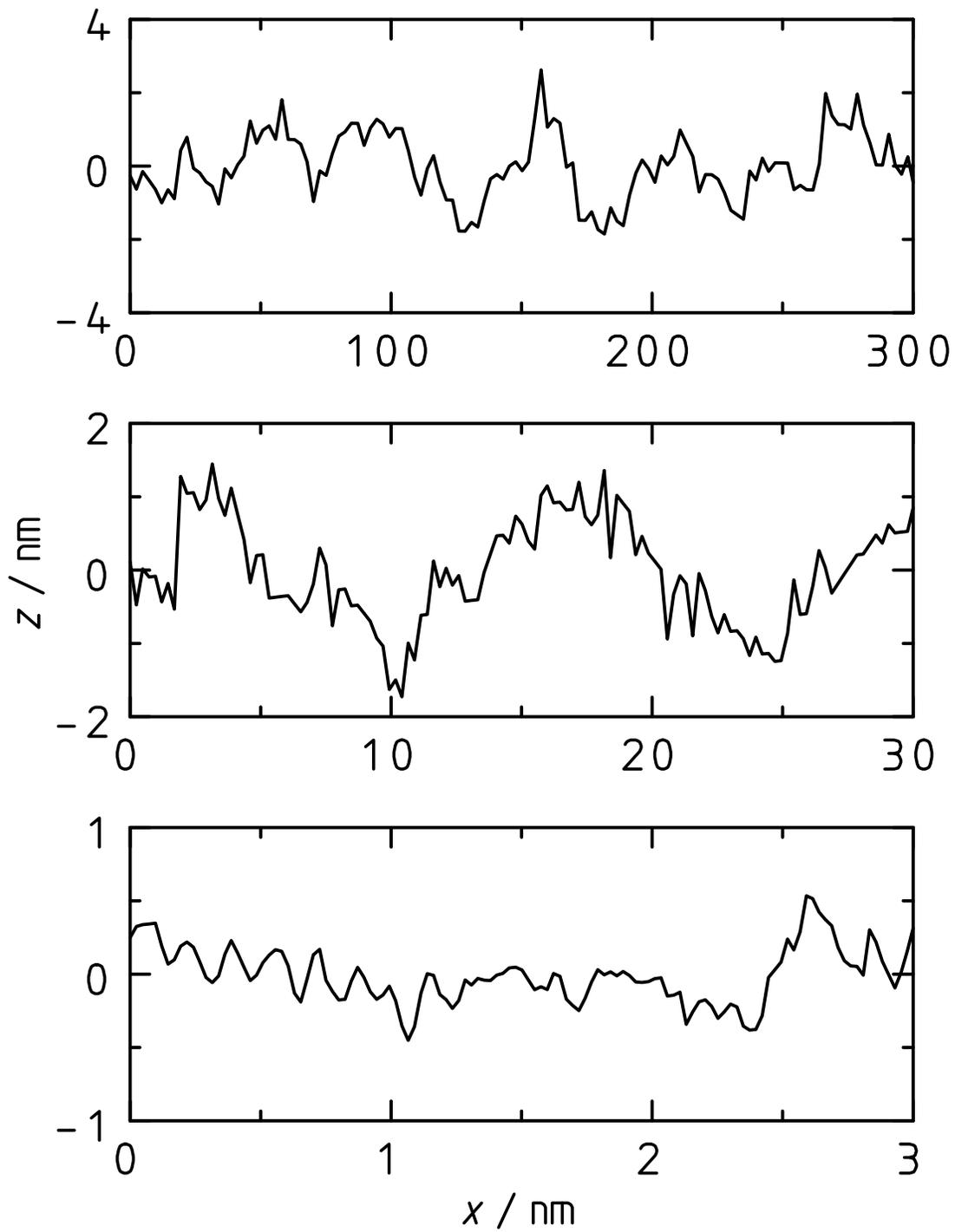
\begin{figure}
\caption{
Three STM cross sections of the surface of sample 3 obtained with 
different resolutions (current = 0.3~nA, bias = 8~V). The data presented 
are obtained from the row data performing only a slope correction. In 
all cases, the height was determined at 128 points along a line segment, 
slightly longer than represented.
}
\label{fig4}
\end{figure}

\begin{figure}
\caption{
Overview of $\sigma(T,x)$ in double logarithmic representations for 
sets A (+) and B ($\times$) in graphs (a) and (b), respectively. 
For the characterization of the samples see Tables I and II. To 
illustrate common as well as differing features, both graphs include a 
sample of the other set, for the sake of distinctness marked by
$\bullet$. Moreover, (a) includes data for two sputtered 
a-Si$_{1-x}$Ni$_x$ films ($\blacktriangle$) from Ref.\ 
{\protect \onlinecite{Isob.etal}}. These samples, here referred to as 
$\alpha$ and $\beta$, have a Ni content of 20.7, and 27.1~at$\%$, 
respectively, as determined by an EDX analysis
{\protect \cite{Isob.etal}}.
}
\label{fig5}
\end{figure}

\begin{figure}
\caption{
$\sigma(T,x)$ versus $T^{1/2}$ representation of the samples closest 
to the MIT for sets A (+) and B ($\times$) in graphs (a) and (b), 
respectively. Both graphs include one sample of the other set 
($\bullet$). The dashed lines give the extrapolations according to Eq.\ 
(2) with $p = 1/2$, obtained from the region 2 -- 30~K.
}
\label{fig6}
\end{figure}

\begin{figure}
\caption{
Overview of the $T$ and $x$ dependence of the logarithmic derivative, 
$w = \mbox{d} \ln \sigma / \mbox{d} \ln T$ for sets A (+) and B 
($\times$) in graphs (a) and (b), respectively. Both graphs include one 
curve of the other set ($\bullet$). The temperature scale $T^{1/2}$ is 
chosen in order to have reasonable resolution also for the low-$T$ part. 
}
\label{fig7}
\end{figure}

\begin{figure}
\caption{
Low-$T$ / low-$w$ part of Fig.\ 7. The data presented were obtained from 
measurements in three different cryostats, see Section IV, where several 
months passed between the investigations. Moreover, in some cases, 
contacts had to be renewed between the measurements.
}
\label{fig8}
\end{figure}

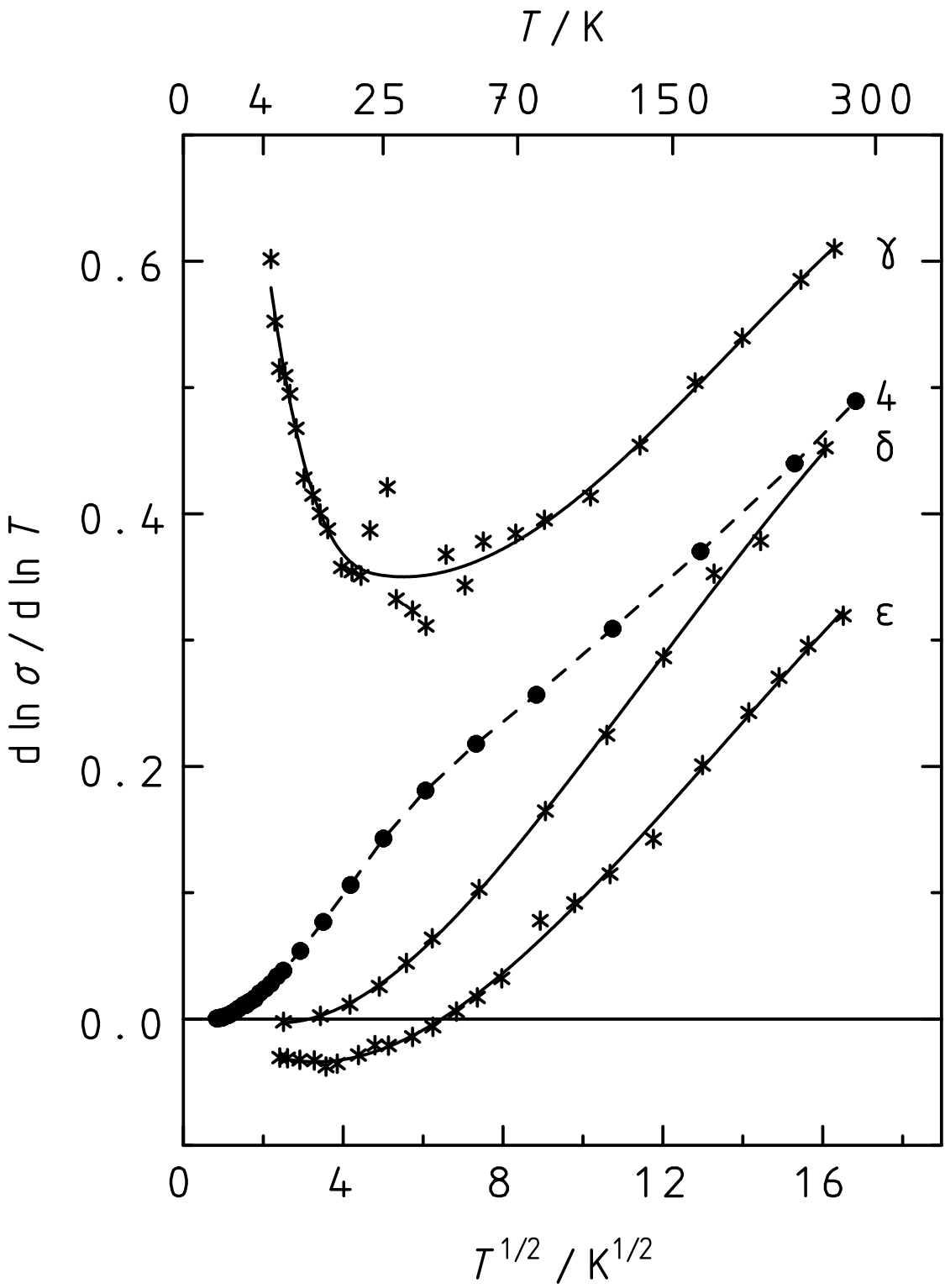
\begin{figure}
\caption{
Comparison of $w(T)$ for the a-Si$_{1-x}$Ni$_x$ sample 4 ($\bullet$) 
with the $w(T)$ for three a-Si$_{1-x}$Cr$_x$ samples ($\ast$), here 
referred to as $\gamma$, $\delta$, and $\varepsilon$. The 
a-Si$_{1-x}$Cr$_x$ curves were obtained from $\sigma(T)$ data, published 
in Figs.\ 1 and 7 of Ref.\ {\protect \onlinecite{Moe.etal.85}}. Samples 
$\gamma$, $\delta$, and $\varepsilon$, have a Cr content of 11, 14, and 
19~at$\%$, respectively, obtained by EDX {\protect \cite{Moe.etal.85}}. 
Calculating $w$, we used $k = 4$ for samples $\gamma$ and $\varepsilon$, 
and $k = 2$ for $\delta$.
}
\label{fig9}
\end{figure}

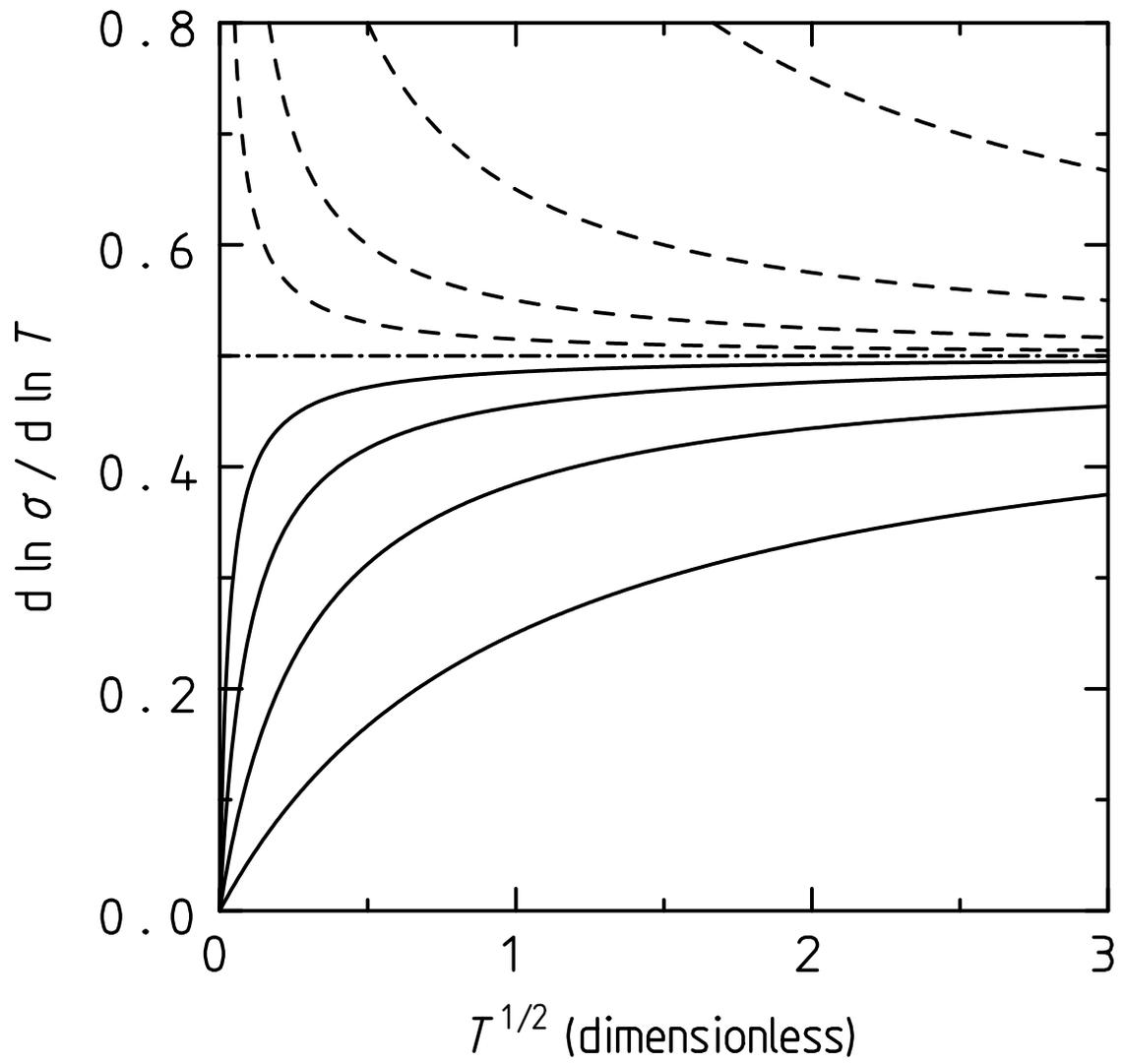
\begin{figure}
\caption{
Qualitative behavior of $w(T,x)$ for a continuous transition according 
to Eq.\ (10). Dashed lines: insulating with 
$T_0(x)^{1/2} = 0.03,\ 0.1,\ 0.3,\ {\rm and}\ 1$; full lines: metallic 
with $a(x) = 0.03,\ 0.1,\ 0.3,\ {\rm and}\ 1$; dashed-dotted line: 
separatrix between insulating and metallic regions.
}
\label{fig10}
\end{figure}

\begin{figure}
\caption{
Characteristic $\sigma(T,x)$ features of a discontinuous transition 
according to Eq.\ (11): $T$ and $x$ dependences of $\sigma$ are shown in 
graphs (a) and (b), respectively. Dashed and full lines represent 
insulating and metallic regions, respectively; the dashed-dotted line 
in (a) denotes the separatrix between these regions. The lines shown 
in (a) were obtained for the parameter values 
$T_0(x)^{1/2} = 0.05,\ 0.15,\ 0.5,\ 1.5$, and 
$a(x) - 1 = 0.2,\ 0.6,\ 1.0,\ 1.4$. Shaded area in (a): example of 
low-$T$ range being inaccessible in a real experiment.
}
\label{fig11}
\end{figure}

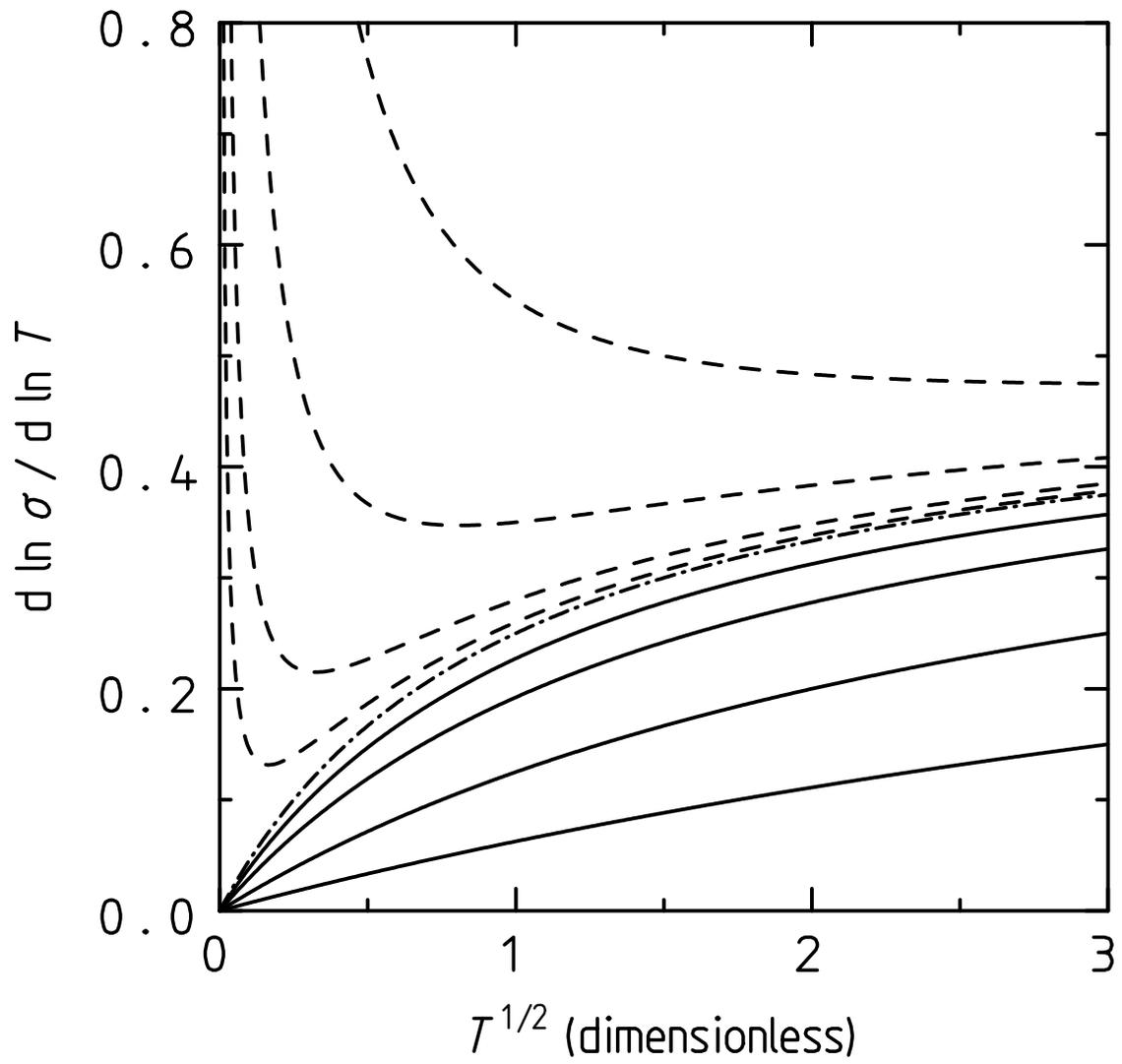
\begin{figure}
\caption{
Qualitative behavior of $w(T,x)$ for a discontinuous transition 
according to Eq.\ (11). Dashed lines: insulating with 
$T_0(x)^{1/2} = 0.02,\ 0.06,\ 0.2,\ {\rm and}\ 0.6$; full lines: 
metallic with $a(x) - 1 = 0.2,\ 0.6,\ 2,\ {\rm and}\ 6$; dashed-dotted 
line: separatrix between insulating and metallic regions.
}
\label{fig12}
\end{figure}


\begin{references}
\bibitem{Ande} P.W.~Anderson,  Phys.\ Rev.\ {\bf 109}, 1492 (1958).
\bibitem{Mott.72} N.F.~Mott, Phil.\ Mag.\ {\bf 26}, 1015 (1972).
\bibitem{Abra.etal} E.~Abrahams, P.W.~Anderson, D.C.~Licciardello, and 
  T.V.~Ramakrishnan, Phys.\ Rev.\ Lett.\ {\bf 42}, 673 (1979).
\bibitem{Fink.83a} A.M.~Finkel'stein, Zh.\ Eksp.\ Teor.\ Fiz.\ 
  {\bf 84}, 168 (1983)  [Sov.\ Phys.\ JETP {\bf 57}, 97 (1983)].
\bibitem{Fink.83b} A.M.~Finkel'stein, Pisma  Zh.\ Eksp.\ Teor.\ Fiz.\ 
  {\bf 37}, 436 (1983) [Sov.\ Phys.\ JETP Lett.\ {\bf 37}, 517 (1983)].
\bibitem{Fink.84} A.M.~Finkel'stein, Zh.\ Eksp.\ Teor.\ Fiz.\ 
  {\bf 86}, 367 (1984) [Sov.\ Phys.\ JETP {\bf 59}, 212 (1984)].
\bibitem{Lee.Rama}  P.A.~Lee and T.V.~Ramakrishnan, Rev.\ Mod.\ Phys.\
  {\bf 57}, 287 (1985). 
\bibitem{Moe.85} A .~M\"obius, J.\ Phys.\ C {\bf 18}, 4639 (1985).
\bibitem{Mott.90} N.F.~Mott, {\it Metal-Insulator Transitions}, Second 
  Edition (Taylor and Francis, London, 1990). 
\bibitem{Kram.McKi} B.~Kramer and A.~MacKinnon, Rep.\ Prog.\ Phys.\ 
  {\bf 56}, 1469 (1993).
\bibitem{Beli.Kirk} D.~Belitz and T.R.~Kirkpatrick, Rev.\ Mod.\ Phys.\ 
  {\bf 66}, 261 (1994).
\bibitem{Edwa.95} P.P.~Edwards, in {\it Perspectives in Solid State 
  Chemistry}, ed.\ K.J.~Rao, (Wiley, New York,1995), p.\ 250.
\bibitem{Paal.etal} M.A.~Paalanen, T.F.~Rosenbaum, G.A.~Thomas, and 
  R.N.~Bhatt, Phys.\ Rev.\ Lett.\ {\bf 48}, 1284 (1982).
\bibitem{Thom.etal.83} G.A.~Thomas, M.~Paalanen, and T.F.~Rosenbaum, 
  Phys.\ Rev.\ B {\bf 27}, 3897 (1983).
\bibitem{Hert.etal} G.~Hertel, D.J.~Bishop, E.G.~Spencer, J.M.~Rowell,
  and R.C.~Dynes, Phys.\ Rev.\ Lett.\ {\bf 50}, 743 (1983).
\bibitem{Peih.etal} Peihua Dai, Youzhu Zhang, and M.P.~Sarachik, Phys.\ 
  Rev.\ Lett.\ {\bf 66}, 1914 (1991).
\bibitem{Stup.etal} H.~Stupp, M.~Hornung, M.~Lakner, O.~Madel, and
  H.~v.~L\"ohneysen, Phys.\ Rev.\ Lett.\ {\bf 71}, 2634 (1993).
\bibitem{Cast} T.G.~Castner,  Phys.\ Rev.\ Lett.\ {\bf 73}, 3600 (1994).
\bibitem{Moe.89} A.~M\"obius, Phys.\ Rev.\ B {\bf 40}, 4194 (1989).
\bibitem{Hirs.etal} M.J.~Hirsch, U.~Thomanschefsky, and D.F.~Holcomb, 
  Phys.\ Rev.\ B {\bf 40}, 4196 (1989).
\bibitem{Thom.Holc} U.~Thomanschefsky and  D.F.~Holcomb, Phys.\ Rev.\ B
  {\bf 45}, 13\,356 (1992).
\bibitem{Rose.etal.94} R.L.~Rosenbaum, M.~Slutzky, A.~M\"obius, and
  D.S.~McLachlan, J.\ Phys.: Cond.\ Matt.\ {\bf 6}, 7977 (1994).
\bibitem{Krav.etal.94} S.V.~Kravchenko, G.V.~Kravchenko, J.E.~Furneaux, 
  V.M.~Pudalov, and M.~D'Iorio, Phys.\ Rev.\ B {\bf 50}, 8039 (1994). 
\bibitem{Krav.etal.95} S.V.~Kravchenko, Whitney E.~Mason, G.E.~Bowker, 
  J.E.~Furneaux, V.M.~Pudalov, and M.~D'Iorio, Phys.\ Rev.\ B {\bf 51}, 
  7038 (1995). 
\bibitem{Krav.etal.96} S.V.~Kravchenko, D.~Simonian, M.P.~Sarachik, 
  Whitney Mason, and J.E.~Furneaux, Phys.\ Rev.\ Lett.\ {\bf 77}, 4938 
  (1996). 
\bibitem{Lubk} G.B.~Lubkin, Physics Today {\bf 50}, No.\ 7, 19 (1997).
\bibitem{Alts.Aron} B.L.~Altshuler and A.G.~Aronov, Zh.\ Eksp.\ Teor.\
  Fiz.\ {\bf 77}, 2028 (1979) [Sov.\ Phys.\ JETP {\bf 50}, 968 (1979)].
\bibitem{News.Pepp} D.J.~Newson and M.~Pepper, J.\ Phys.\ C {\bf 19},
  3983 (1986).
\bibitem{Moe.etal.83} A.~M\"obius, D.~Elefant, A.~Heinrich, 
  R.~M\"uller, J.~Schumann, H.~Vinzelberg,  and G.~Zies, J.\ Phys.\ C
  {\bf 16}, 6491 (1983). 
\bibitem{Moe.etal.85} A.~M\"obius, H.~Vinzelberg, C.~Gladun, 
  A.~Heinrich, D.~Elefant, J.~Schumann, and G.~Zies, J.\ Phys.\ C 
  {\bf 18}, 3337 (1985). 
\bibitem{Moe.87} A.~M\"obius, phys.\ stat.\ sol.\ (b) {\bf 144},
  759 (1987).
\bibitem{Moe.90a} A.~M\"obius, Z.\ Phys.\ B {\bf 79}, 265 (1990).
\bibitem{Moe.90b} A.~M\"obius, Z.\ Phys.\ B {\bf 80}, 213 (1990).
\bibitem{Phasediag} {\it Binary Alloy Phase Diagrams}, ed.\ in chief:
  T.B. Massalski, second edition, (ASM International, Materials Park,
  1992).
\bibitem{Nish.etal} N.~Nishida, M.~Yamaguchi, T.~Furubayashi,
  K.~Morigaki, H.~Ishimoto, and K.~Ono, Solid State Commun.\ {\bf 44},
  305 (1982), and refs.\ therein.
\bibitem{Dods.etal} B.W.~Dodson, W.L.~McMillan, J.M.~Mochel, and 
  R.C.~Dynes, Phys.\ Rev.\ Lett.\ {\bf 46}, 46 (1981).
\bibitem{Ohri.glass} M.~Ohring, {\it The Materials Science of Thin 
  Films}, (Academic Press, Boston, 1992), p.\ 235.
\bibitem{Olow.etal} J.O.~Olowolafe, M.A.~Nicolet, and J.W.~Mayer, J.\ 
  Appl.\ Phys.\ {\bf 47}, 5182 (1976). 
\bibitem{Colg.etal} E.G.~Colgan, B.Y.~Tsaur, and J.W.~Mayer, Appl.\ 
  Phys.\ Lett.\ {\bf 37}, 938 (1980). 
\bibitem{Sobe.Zies} G.~Sobe and G.~Zies, private communication (1984).
\bibitem{metal_silic} {\it Properties of Metal Silicides}, eds.\ 
  K.~Maex and M.~van~Rossum, EMIS Datareviews Series No.\ 14, (INSPEC, 
  London, 1995).
\bibitem{Abke.etal.92a} K.M.~Abkemeier, C.J.~Adkins, R.~Asal, and 
  E.A.~Davis, J.\ Phys.: Cond.\ Matt.\ {\bf 4}, 9113 (1992). 
\bibitem{Abke.etal.92b} K.M.~Abkemeier, C.J.~Adkins, R.~Asal, and
  E.A.~Davis, Phil.\ Mag.\ B {\bf 65}, 675 (1992).
\bibitem{Damm.etal} U.~Dammer, C.J.~Adkins, R.~Asal, and E.A.~Davis, 
  J.\ Non-Cryst.\ Sol.\ {\bf 164--166}, 501 (1993). 
\bibitem{Damm} U.~Dammer, {\it The Metal-Insulator Transition in 
  Amorphous Silicon-Nickel Alloys}, Master of Philosophy Thesis, 
  University of Cambridge (1992).
\bibitem{Coll1} M.M.~Collver, Solid State Commun.\ {\bf 23}, 333 (1977).
\bibitem{Coll2} M.M.~Collver, Appl.\ Phys.\ Lett.\ {\bf 32}, 574 (1978).
\bibitem{Sega.etal} C.~Segal, A.~Gladkikh, M.~Pilosof, H.~Behar,
  M.~Witcomb, and R.~Rosenbaum, J.\ Phys.: Condens.\ Matter {\bf 10},
  123 (1998).
\bibitem{Belu.etal} A.~Belu-Marian, M.D.~Serbanescu, R.~Manaila, 
  E.~Ivanov, O.~Malis, A.~Devenyi, Thin Solid Films {\bf 259}, 105 
  (1995).
\bibitem{Isob.etal} A.~Isobe, M.~Yamada, and K.~Tanaka, J.\ Phys.\ Soc.\
  Jap.\ {\bf 66}, 2103 (1997).
\bibitem{Edwa.etal} A.M.~Edwards, M.C.~Fairbanks, A.~Singh, R.J.~Newport,
  and S.J.~Gurman, Physica B {\bf 158}, 600 (1989).
\bibitem{Davi.etal} E.A.~Davis, S.C.~Bayliss, R.~Asal, and M.~Manssor, 
  J.\ Non-Cryst.\ Sol.\ {\bf 114}, 465 (1989).
\bibitem{Asal.etal} R.~Asal, S.C.~Bayliss, and E.A.~Davis, J.\ 
  Non-Cryst.\ Sol.\ {\bf 137\&138}, 931 (1991).
\bibitem{Gheo.etal} A.~Gheorghiu, C.~S\'en\'emaud, R.~Asal, and 
  E.A.~Davis, J.\ Non-Cryst.\ Sol.\ {\bf 182}, 293 (1995). 
\bibitem{Rose.etal.97a} R.~Rosenbaum, A.~Heines, A.~Palevski,
  M.~Karpovski, A.~Gladkikh, M.~Pilosof, A.J.~Daneshvar, M.R.~Graham,
  T.~Wright, J.T.~Nicholls, C.J.~Adkins, M.~Witcomb, V.~Prozesky, 
  W.~Przybylowicz, and R.~Pretorius, J.\ Phys.: Condens.\ Matter {\bf 9}, 
  5395 (1997).
\bibitem{Rose.etal.97b} R.~Rosenbaum, A.~Heines, M.~Karpovski,
  M.~Pilosof, and M.~Witcomb, J.\ Phys.: Condens.\ Matter {\bf 9}, 5413
  (1997).
\bibitem{Ohri} M.~Ohring, {\it The Materials Science of Thin Films},
  (Academic Press, Boston, 1992), p.\ 294.
\bibitem{EDX} A.P.~Mackenzie, Rep.\ Prog.\ Phys.\ {\bf 56}, 557 (1993).
\bibitem{Hoff} V.~Hoffmann, H.-J.~Uhlemann, F.~Pr\"a{\ss}ler, K.~Wetzig,
  and D.~Birus, Fresenius J.\ Anal.\ Chem.\ {\bf 355}, 826 (1996).
\bibitem{Raap.etal} M.B.~Fernandez van Raap, M.J.~Regan, and
  A.~Bienenstock, J.\ Non-Cryst.\ Sol.\ {\bf 191}, 155 (1995).
\bibitem{Bara.Stan} A.-L.~Barabasi and H.E.~Stanley, 'Fractal
  Concepts in Surface Growth', Cambridge University Press, 1995.
\bibitem{Bala.Zang} G.S.~Bales and A.~Zangwill, J.\ Vac.\ Sci.\ 
  Technol.\ A {\bf 9}, 145 (1991).
\bibitem{Kran.Lodd} H.~van Kranenburg and C.~Lodder, Mat.\ Sc.\ and 
  Eng.\ {\bf R11}, 295 (1994). 
\bibitem{Frie.etal} Jonathan R.~Friedman, Youzhu Zhang, Peihua Dai, and 
  M.P.~Sarachik, Phys.\ Rev.\ B {\bf 53}, 9528 (1996).
\bibitem{Zabr.Zino} A.G.~Zabrodskii and K.N.~Zinov'eva, Zh.\ Eksp.\
  Teor.\ Fiz.\ {\bf 86}, 727 (1984) [Sov.\ Phys.\ JETP {\bf 59}, 425 
  (1984)].
\bibitem{Moe.88} A.~M\"obius, J.\ Phys.\ C {\bf 21}, 2789 (1988).
\bibitem{Wrig.etal} T.~Wright, B.~Popescu, C.J.~Adkins, and E.A.~Davis,
  J.\ Phys.: Condens.\ Matter {\bf 8}, 6737 (1996).
\bibitem{Glad.etal} C.~Gladun, A.~Heinrich, F.~Lange, J.~Schumann, and
  H.~Vinzelberg, Thin Solid Films {\bf 125}, 101 (1985).
\bibitem{Albe.McLa} A.~Albers and D.S.~McLachlan, Czech.\ J.\ Phys.\
  {\bf 46}, Suppl.\ S2, 753 (1996).
\bibitem{Rose.etal.81} T.F.~Rosenbaum, K.~Andres, G.A.~Thomas, and 
  P.A.~Lee, Phys.\ Rev.\ Lett.\ {\bf 46}, 568 (1981).
\bibitem{Thom.etal.82} G.A.~Thomas, A.~Kawabata, Y.~Ootuka, 
  S.~Katsumoto, S.~Kobayashi, and W.~Sasaki, Phys.\ Rev.\ B {\bf 26}, 
  2113 (1982).
\bibitem{Mott.Davi.79} N.F.~Mott and E.A.~Davis, {\it Electronic 
  Processes in Non-Crystalline Materials}, Second Edition (Clarendon, 
  Oxford, 1979).
\bibitem{Mott.56} N.F.~Mott, Can.\ J.\ Phys.\ {\bf 34}, 1356 (1956).
\bibitem{Edwa.Sien.78} P.P.~Edwards and M.~Sienko, Phys.\ Rev.\ B 
  {\bf 17}, 2575 (1978). 
\bibitem{Edwa.Sien.83} P.P.~Edwards and M.~Sienko, Int.\ Rev.\ Phys.\ 
  Chem.\ {\bf 3}, 83 (1983).
\bibitem{Shaf.etal} W.N.~Shafarman, D.W.~Koon, and T.G.~Castner, Phys.\ 
  Rev.\ B {\bf 40}, 1216 (1989).
\bibitem{SiCrscale} In Refs.\ 
  \protect{\onlinecite{Moe.etal.83,Moe.etal.85}}, increase of $w$ 
  with decreasing $\sigma$ is presented for Si$_{1-x}$Cr$_x$. Due 
  to $\mbox{d} \sigma / \mbox{d} T > 0$, this corresponds to 
  $\mbox{d} w / \mbox{d} T < 0$.
\bibitem{Yaki.etal} A.I.~Yakimov, A.V.~Dvurechenskii, and C.J.~Adkins, 
  phys.\ stat.\ sol.\ (b) {\bf 205}, 299 (1998).
\bibitem{Vinz.etal} H.~Vinzelberg, A.~Heinrich, C.~Gladun, and 
  D.~Elefant, Phil.\ Mag.\ B {\bf 65}, 651 (1992).
\bibitem{Mott.68} N.F.~Mott, J.\ Non-Cryst.\ Sol.\ {\bf 1}, 1 (1968).
\bibitem{Efro.Shkl} A.L.~Efros and B.I.~Shklovskii, J.\ Phys.\ C
  {\bf 8}, L49 (1975).
\bibitem{Adki} C.J.~Adkins, J. Phys.: Condens.\ Matter {\bf 1}, 1253
  (1989).
\bibitem{Poll.98} M.~Pollak, phys.\ stat.\ sol.\ (b) {\bf 205}, 35 
  (1998).
\bibitem{Chit.Kotl} R.~Chitra and G.~Kotliar, cond-mat/9903180.
\end{references}
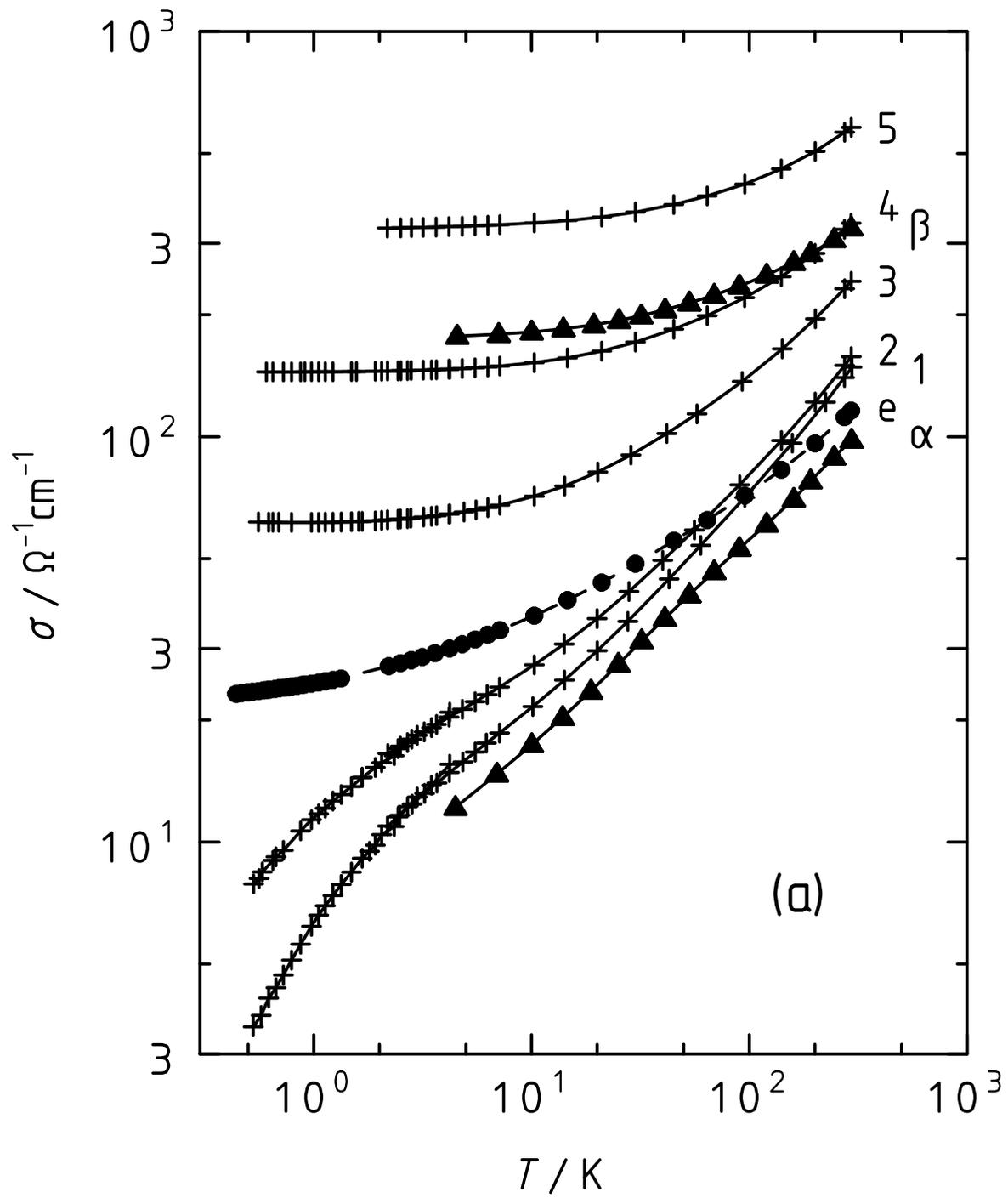
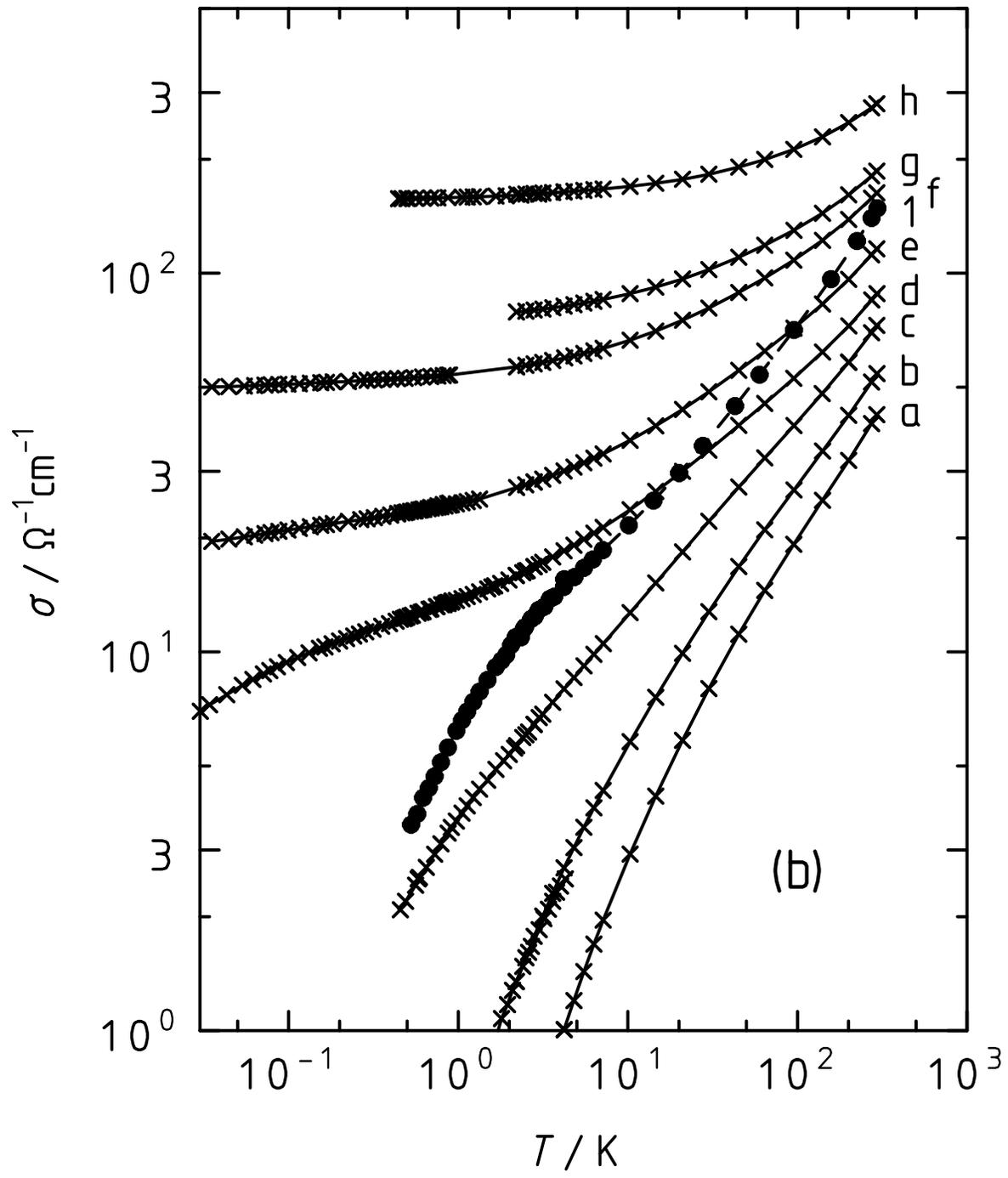
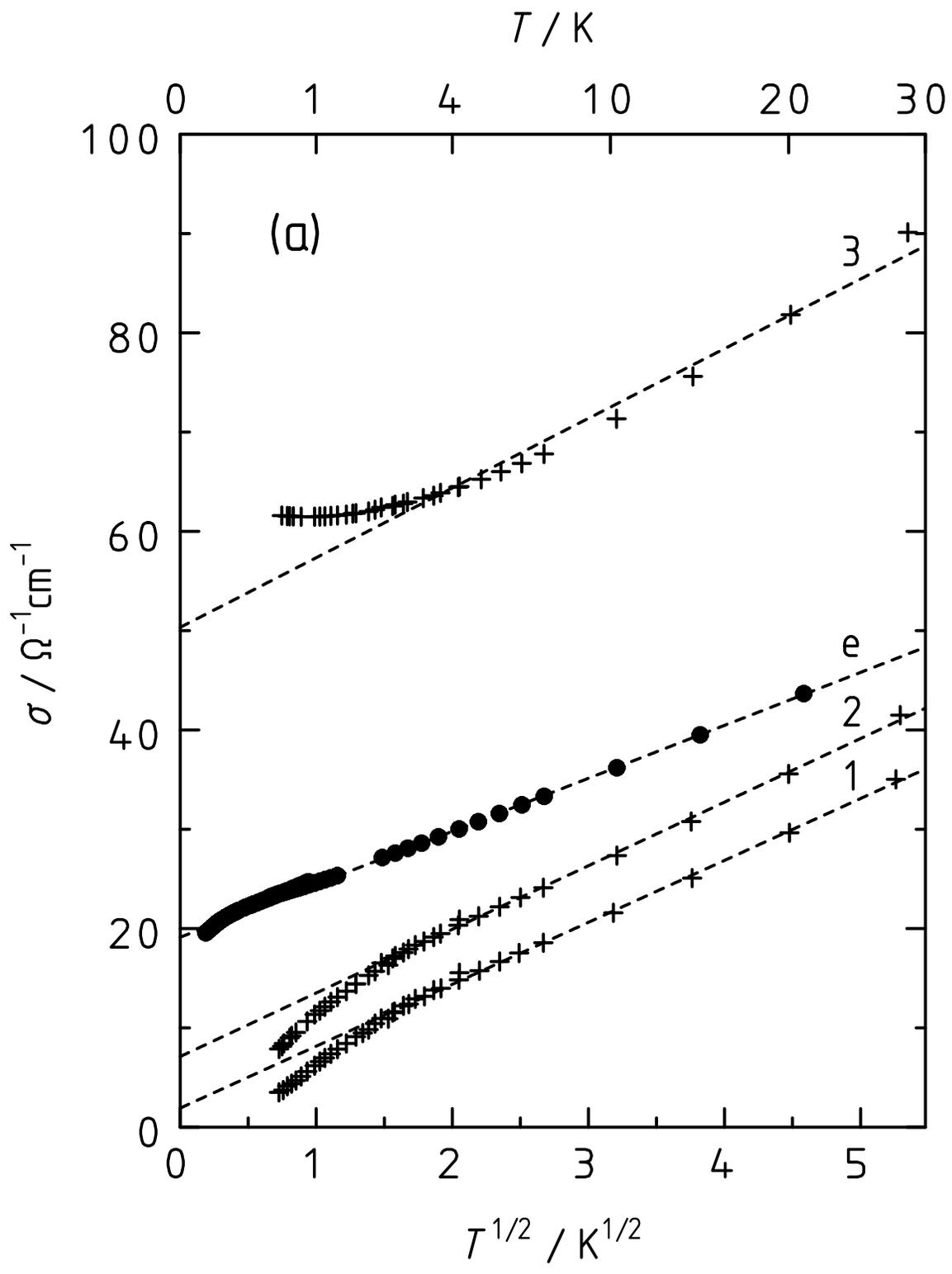
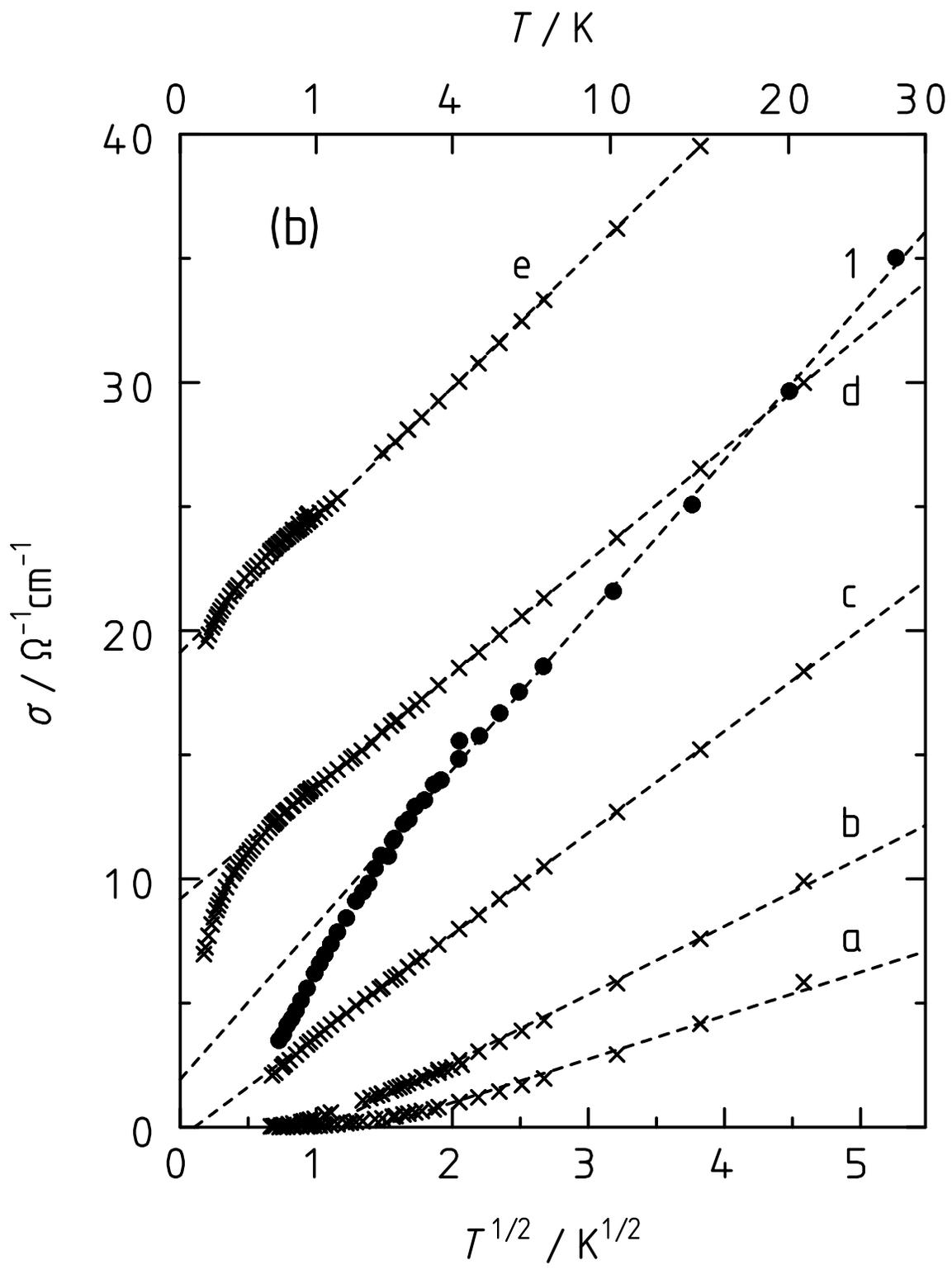
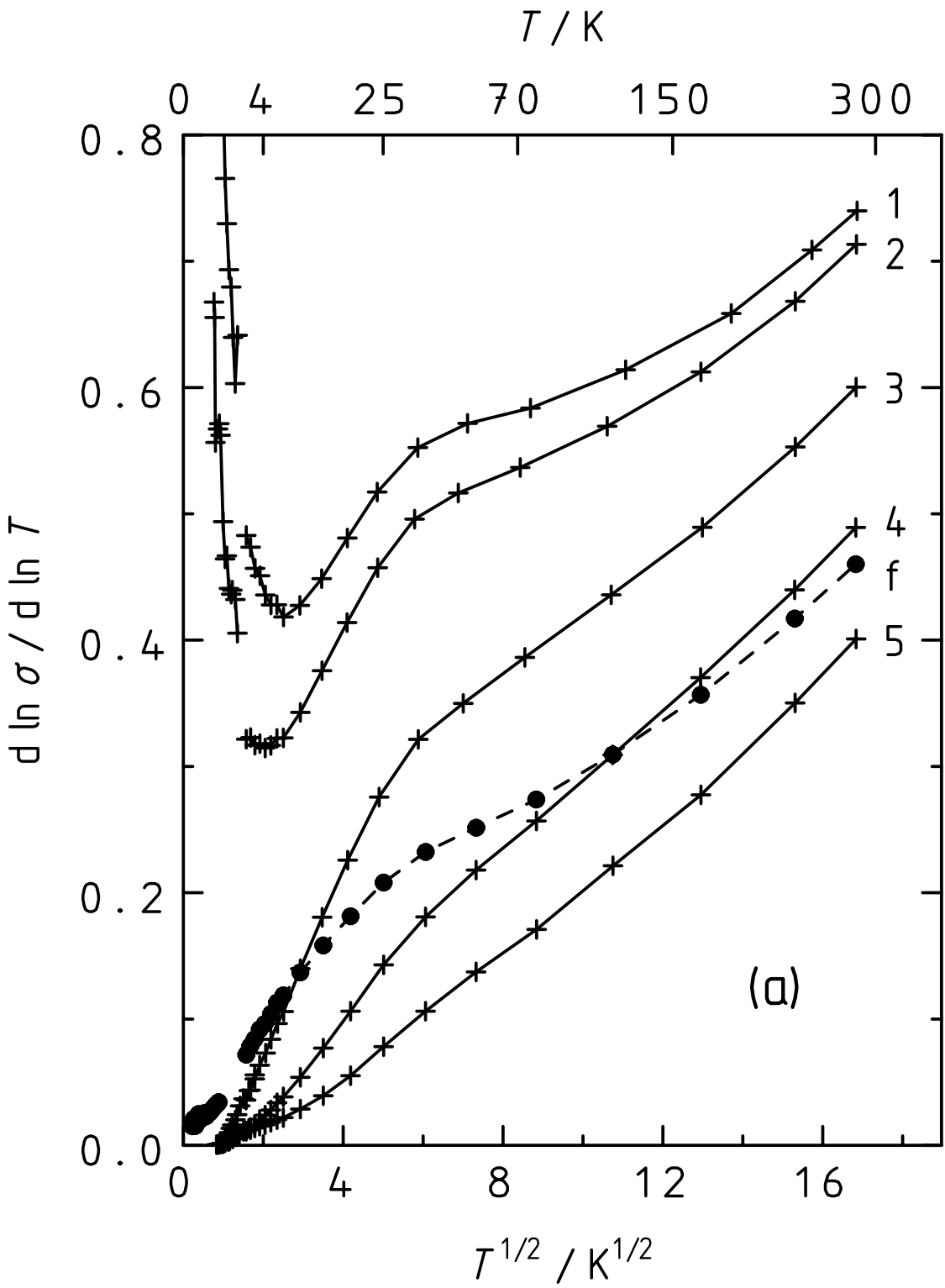
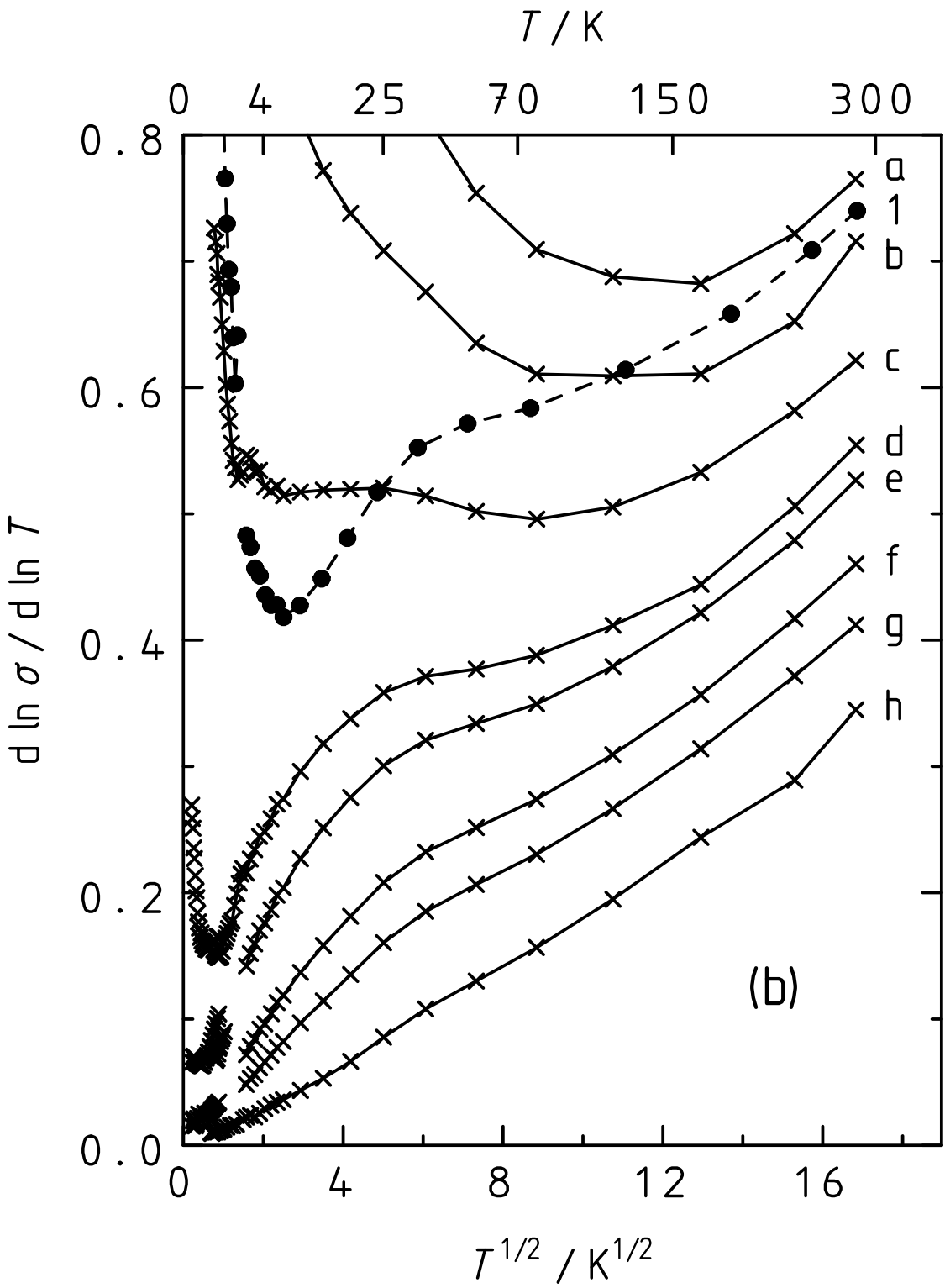
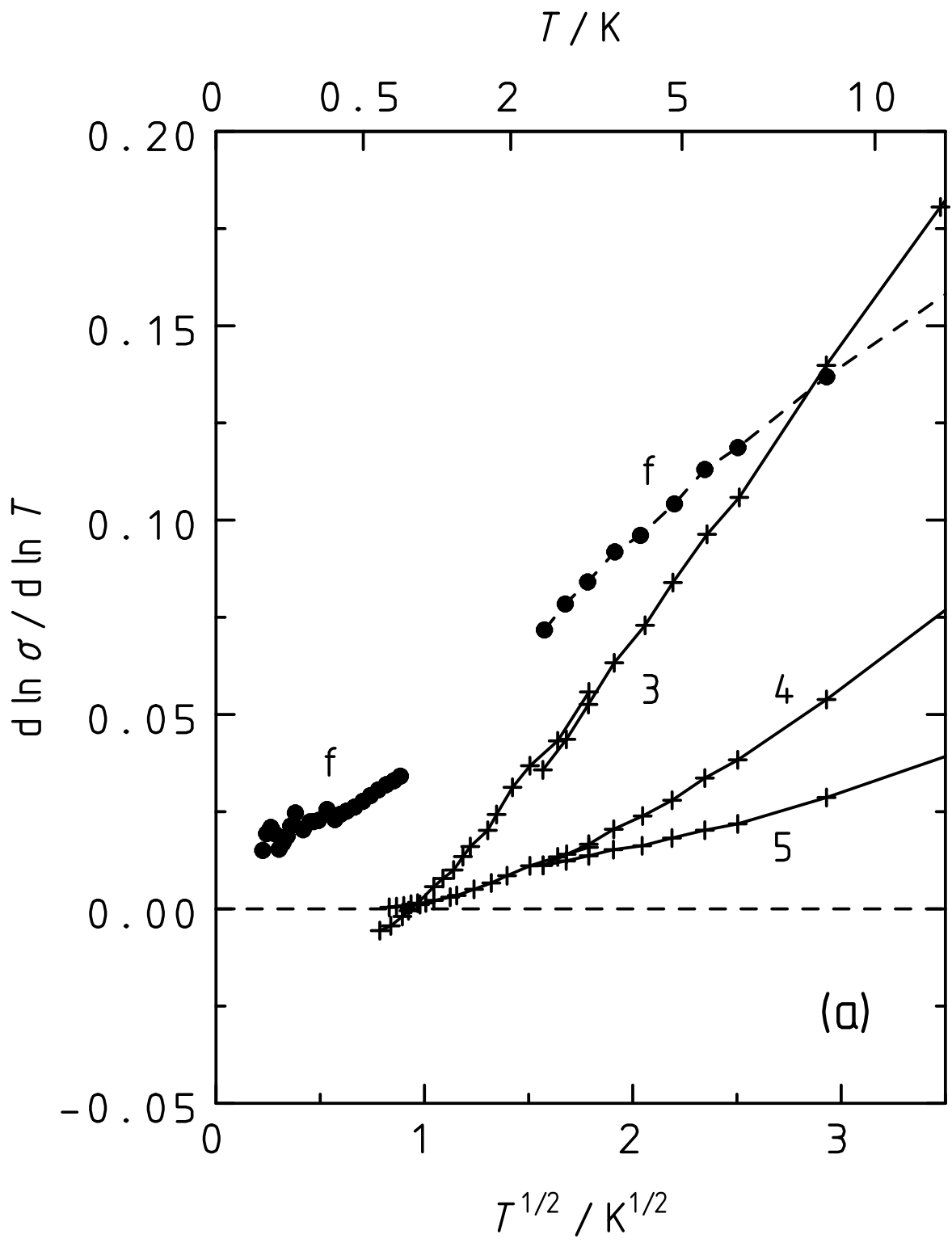
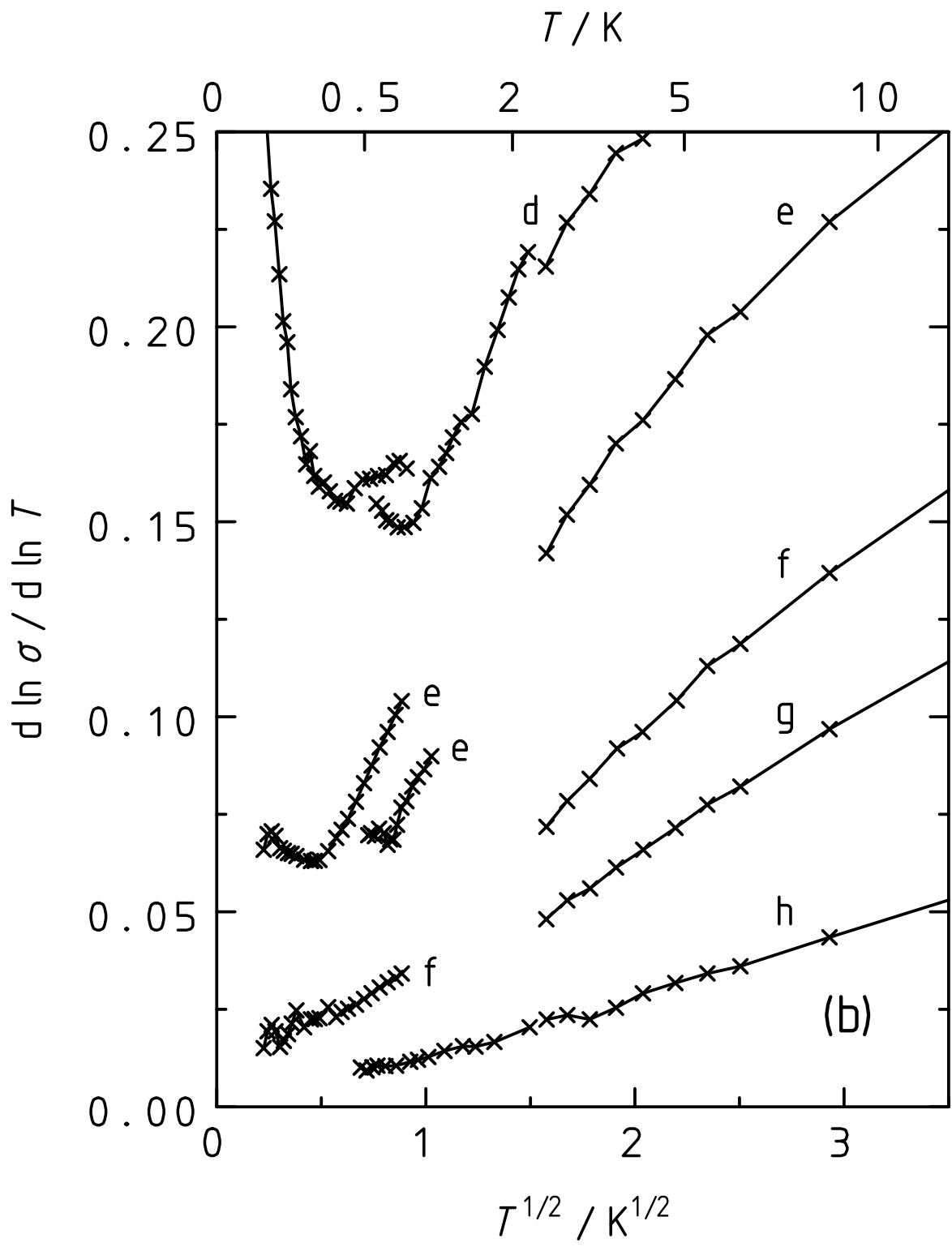
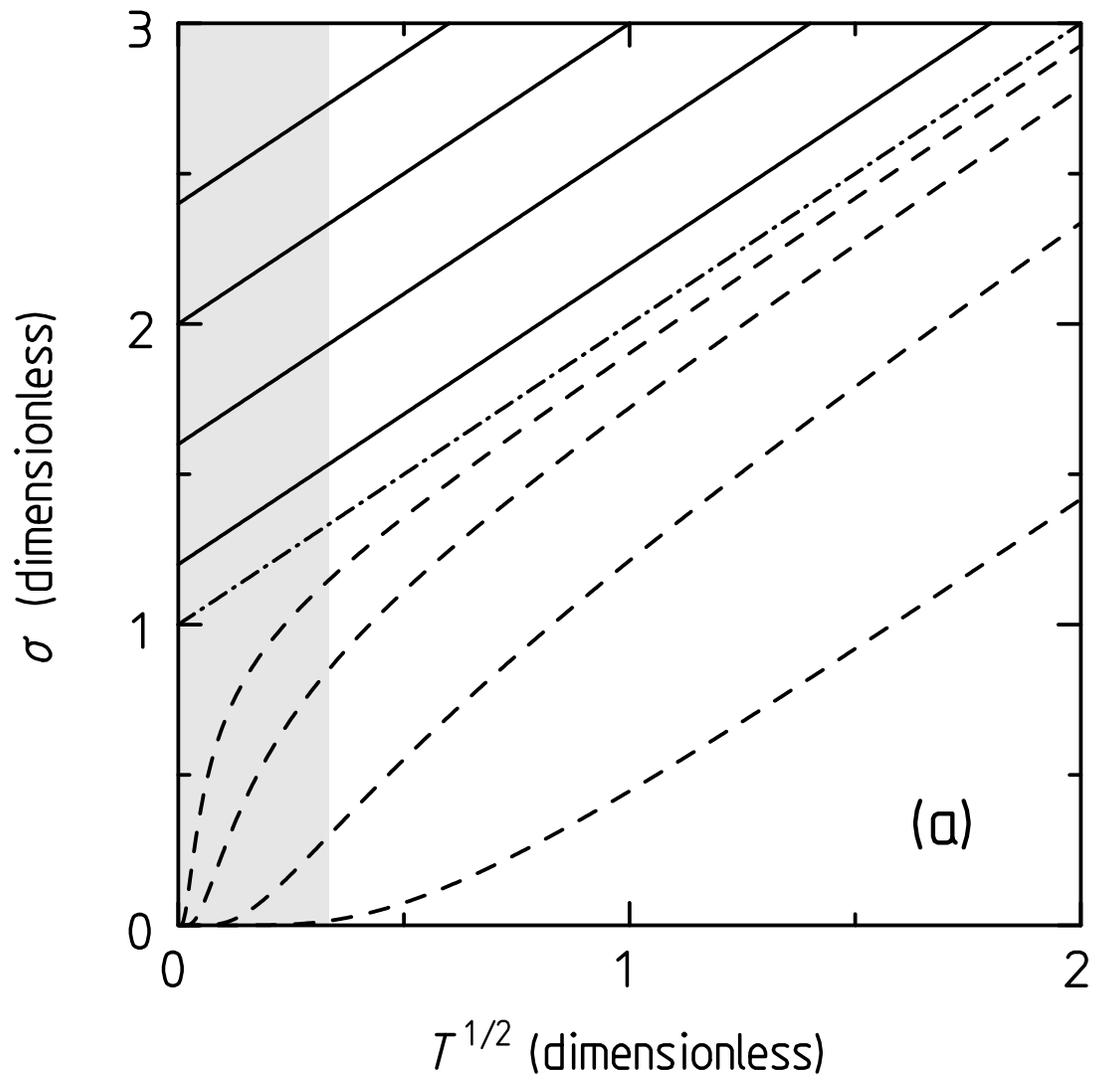
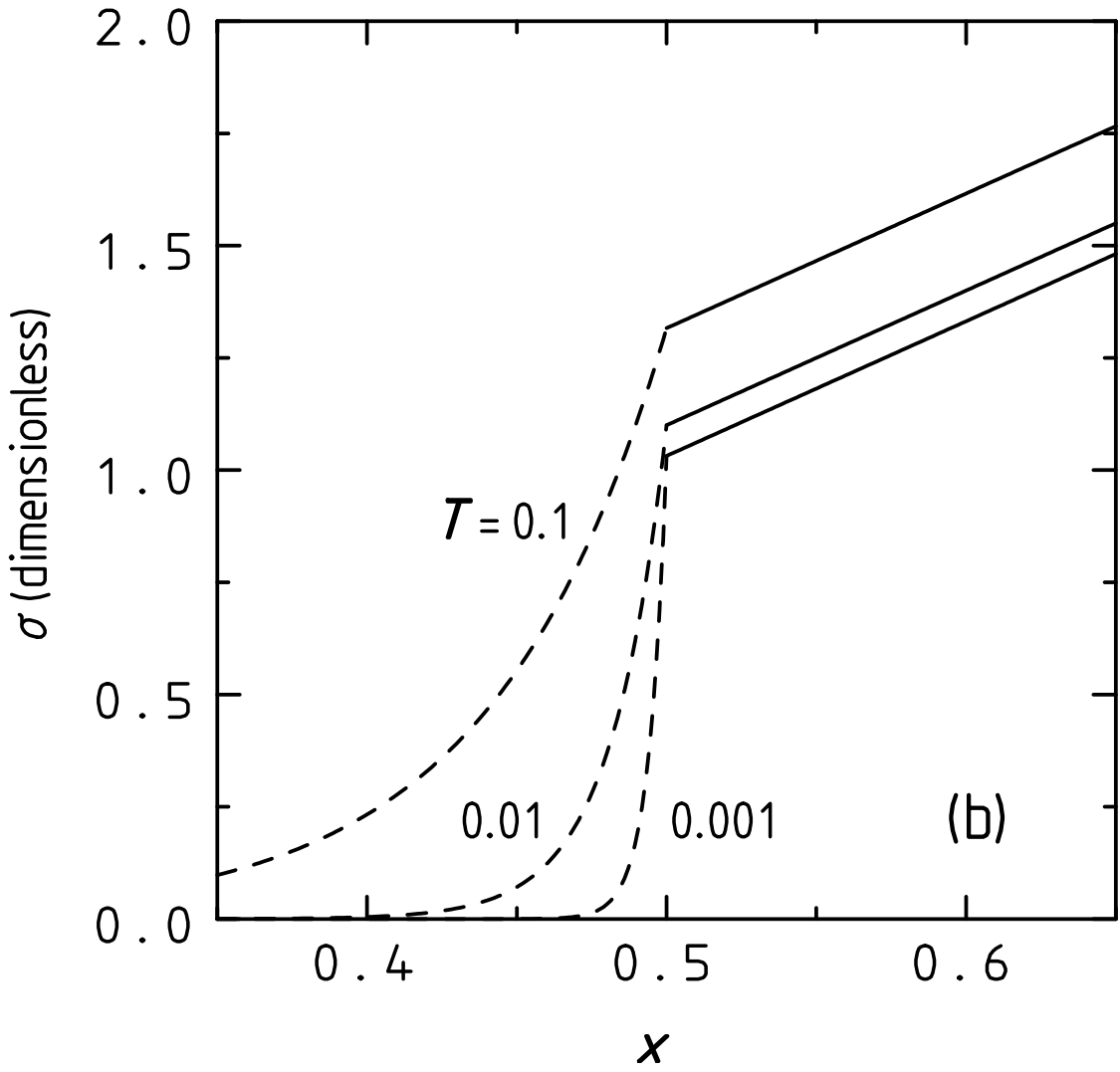
\end{document}